\documentclass[manuscript]{aastex61}
\pdfoutput=1
\usepackage{amsmath,amstext}
\usepackage[T1]{fontenc}
\usepackage[figure,figure*]{hypcap}
\usepackage{tikz}
\usetikzlibrary{shapes,arrows}

\newcommand{\logg} {\log \textsl{\textrm{g}}}
\newcommand{\Te} {T_{\rm eff}}

\newcommand{\msun} {M_\odot}

\newcommand{\K}{{\rm K}}
\newcommand\gta{\lower 0.5ex\hbox{$\buildrel > \over \sim\ $}} 
\newcommand\lta{\lower 0.5ex\hbox{$\buildrel < \over \sim\ $}} 

\newcommand{\logH}{\log {\rm H}/{\rm He}}

\newcommand{\gaia}{\textit{Gaia~}}
\newcommand{\sn}{{\rm S/N}}
\newcommand{\logCa}{\log {\rm Ca}/{\rm He}}
\newcommand{\nHH}{{\rm H}/{\rm He}}

\newcolumntype{L}{>{$}c<{$}}

\shorttitle{Analysis of DB WDs from SDSS and Gaia}
\shortauthors{Genest-Beaulieu \& Bergeron}

\begin{document}

\title{A Photometric and Spectroscopic Investigation of the DB White
  Dwarf Population using SDSS and \textit{Gaia} Data}

\author{C. Genest-Beaulieu}
\affiliation{D\'epartement de Physique, Universit\'e de Montr\'eal, Montr\'eal, 
QC H3C 3J7, Canada; genest@astro.umontreal.ca, bergeron@astro.umontreal.ca.}
\author{P. Bergeron}
\affiliation{D\'epartement de Physique, Universit\'e de Montr\'eal, Montr\'eal, 
QC H3C 3J7, Canada; genest@astro.umontreal.ca, bergeron@astro.umontreal.ca.}

\begin{abstract}

We present a comprehensive analysis of DB white dwarfs drawn from the
Sloan Digital Sky Survey, based on model fits to $ugriz$ photometry
and medium resolution spectroscopy from the SDSS. We also take
advantage of the exquisite trigonometric parallax measurements
recently obtained by the {\it Gaia} mission. Using the so-called
photometric and spectroscopic techniques, we measure the atmospheric
and physical parameters of each object in our sample ($\Te$, $\logg$,
$\nHH$, Ca/He, $R$, $M$), and compare the values obtained from both
techniques in order to assess the precision and accuracy of each
method. We then explore in great detail the surface gravity, stellar
mass, and hydrogen abundance distributions of DB white dwarfs as a
function of effective temperature. We present some clear evidence for
a large population of unresolved double degenerate binaries composed
of DB+DB and even DB+DA white dwarfs. In the light of our results, we
finally discuss the spectral evolution of DB white dwarfs, in
particular the evolution of the DB-to-DA ratio as a function of $\Te$,
and we revisit the question of the origin of hydrogen in DBA white
dwarfs.

\end{abstract}


\section{Introduction}

The Sloan Digital Sky Survey (SDSS) has dramatically changed our view
of white dwarf stars by not only increasing the number of known
degenerates by a factor of more than 10, but also by providing a set
of homogeneous spectroscopic and photometric observations
\citep{Abazajian2003}. More recently, another major milestone has been
achieved by the {\it Gaia} mission \citep{gaia2}, which provided
accurate trigonometric parallax measurements for 260,000 white dwarfs,
and white dwarf candidates \citep{Gentile2019}. These accurate
distances, coupled with large photometric surveys such as SDSS, {\it
  Gaia}, or Pan-STARRS, open an all new window on the measurement of
white dwarf parameters
\citep{Gentile2019,Tremblay2019,GBB19,Bergeron2019}. In
\citet[][hereafter GBB19]{GBB19}, we presented a detailed comparison
of white dwarf parameters obtained with the so-called spectroscopic
and photometric techniques for both DA and DB stars (see also
\citealt{Genest2014} and \citealt{Tremblay2019}). In this paper, we
want to focus more specifically on the origin and formation of DB
white dwarfs.

While a countless number of spectroscopic analyses of DA white dwarfs
has been published in the literature (\citealt{BSL92},
\citealt{LBH05}, \citealt{Koester2009}, and \citealt{Gianninas2011},
just to name a few), only a few are available for DB stars
\citep{Eisenstein2006,Voss2007,Bergeron2011,Koester2015,Rolland2018}. More
importantly, while the atmospheric parameters for DA white dwarfs
agree generally well between these various analyses, those of DB stars
show larger discrepancies, which can probably be traced back to
differences in model atmospheres and fitting techniques. For instance,
the treatment of van der Waals broadening remains one of the largest
source of uncertainty at low effective temperatures ($\Te\lesssim
16,000$ K; GBB19 and references therein). Also of importance
is the assumed convective efficiency at high temperatures ($\Te\sim
25,000$ K), or even the validity of the mixing-length theory used so
far in all model atmosphere calculations for DB
stars. \citet{Cukanovaite2018} indeed showed that 3D hydrodynamical
effects, similar to those found in the context of DA stars
\citep{Tremblay2013,TremblayIV}, should be equally important for DB
white dwarfs. Hence, despite all these efforts, our understanding of
the effective temperature, stellar mass, and hydrogen abundance
distributions of DB white dwarfs remains sketchy at best. For
instance, there are still unanswered questions regarding the mass
distribution of DB versus DBA white dwarfs, or the reality of low- and
high-mass DB white dwarfs, or even the existence of unresolved double
degenerate binaries among the DB population.

Many open questions also remain concerning the origin of DB white
dwarfs. There is now little doubt that most DB white dwarfs have
evolved from the transformation of DA stars through a process referred
to as convective dilution, where the thin hydrogen surface layer
($M_{\rm H} \sim 10^{-16}-10^{-14} \msun$) of the DA progenitor is
gradually eroded and thoroughly mixed with the underlying helium
convection zone. However, the details of this mixing process, and in
particular the temperature at which it takes place, remain poorly
understood \citep{MacDonald1991,Rolland2018}. There is also the
question of the existence of the DB-gap, a region between $\Te\sim
45,000$ K and 30,000 K originally believed to be devoid of
helium-atmosphere white dwarfs \citep{Liebert1986,Fontaine1987}. It is
this particular feature that actually led to the interpretation of the
DA-to-DB transition at the red edge of the gap. Now thanks to the
SDSS, this gap has been partially filled by hot DB stars
\citep{Eisenstein2006,Koester2015}, but a strong deficiency of
helium-atmosphere white dwarfs still remains in this temperature
range, and the exact fraction is uncertain due, once again, to
inaccuracies in the temperature scale of hot DB stars.

Important insight can also be gained from a careful determination of
the ratio of DB to DA stars as a function of effective temperature,
ideally in a volume-limited sample to avoid all possible selection
biases. Since the convective dilution process is a strongly-dependent
function of the thickness of the hydrogen layer, one could in
principle map the hydrogen layer mass ($M_{\rm H}$) in DA stars that
turned into DB white dwarfs, by carefully comparing their respective
luminosity functions. For instance, \citet[][see their Figure
  24]{Bergeron2011} used the DA and DB white dwarfs identified in the
Palomar-Green (PG) survey to show that the DA-to-DB transition
occurred for most objects around $\Te\sim 20,000$ K, instead of the
canonical value of 30,000 K, an estimate originally based of the
location of the red edge of the DB gap. This result led the authors to
suggest that a fraction of DB stars may have preserved a helium-rich
atmosphere throughout their lifetime, an interpretation certainly
supported by the existence of hot, helium-atmosphere white dwarfs in
the DB gap.

Another topic of importance is related to the presence of hydrogen in
DB white dwarfs --- the DBA stars. Hydrogen is detected at the
photosphere of a significant fraction of DB stars --- mostly through
spectroscopic observations at H$\alpha$ ---, although the exact
fraction varies from study to study depending on the quality of the
observations, and most importantly the signal-to-noise ratio
(S/N). For instance, \citet{Bergeron2011} found that 44\% of their
sample of 108 objects showed hydrogen, but further spectroscopic
observations at H$\alpha$ of the same sample by \citet{Rolland2018}
increased this ratio to 63\%. On the other hand, \citet{Koester2015}
estimated that this fraction could be as high as 75\% based on the
best spectroscopic data in the SDSS.

An important controversy in the literature also has to do with the
origin of hydrogen in these DBA white dwarfs. One possible explanation
is that hydrogen has a residual origin, resulting from the convective
dilution of the thin hydrogen layer with the more massive helium
convection zone. However, the total mass of hydrogen within the mixed
H/He convection zone, inferred from the observed hydrogen abundance at
the photosphere, is so large --- of the order of $M_{\rm
  H}=10^{-12}~\msun$ --- that a DA progenitor with such a massive
hydrogen layer would have never mixed in the first place. Or put
differently, the thickness of the hydrogen layer required for the
convective dilution process to occur would yield photospheric hydrogen
abundances that are orders of magnitudes smaller than those observed
in DBA white dwarfs.

One possible solution to this problem is to have an external source of
hydrogen that would increase its photospheric abundance significantly,
\textit{assuming mixing has already occurred}. Many external sources
have been proposed in the literature, including accretion from the
interstellar medium \citep{MacDonald1991}, from comets
\citep{Veras2014}, or even from disrupted planets
\citep{Raddi2015,Gentile2017}. One problem with this interpretation,
however, is that the average hydrogen accretion rate required to
account for the observed abundances in DBA white dwarfs, would build
over time a hydrogen layer at the surface of the DA progenitor thick
enough, that such a DA star would never undergo the DA-to-DB
transition \citep{Rolland2018}. Also, one would have to explain the
existence of DB stars with no detectable traces of hydrogen, in
particular at low temperatures where small traces of hydrogen ($\logH
\sim -6$) can be easily detected. Nevertheless, there are obvious
cases of DBA stars with extremely large abundances of hydrogen and
metals (SDSS J124231.07+522626.6, GD 16, GD 17, GD 61, and GD 362; see
\citealt{Gentile2017} and references therein), for which the
interpretation in terms of accretion of water-rich asteroid debris
cannot be questioned. \citet{Gentile2017} even discuss a possible
correlation between the presence of hydrogen and metals in DBA white
dwarfs, suggesting that some fraction of the hydrogen detected in
many, perhaps most, helium-atmosphere white dwarfs is accreted
alongside metal pollutants.

In order to shed some light on several of the issues discussed above,
we present in this paper a thorough photometric and spectroscopic
analysis of the DB white dwarfs identified in the SDSS. We first
describe in Section \ref{sect2:sample} the DB white dwarf sample drawn
from the SDSS database, including the spectroscopic, photometric, and
astrometric observations, which will be analyzed using the theoretical
framework outlined in Section \ref{sect2:theory}. We also explore at
length in Section \ref{sect2:error} the error budget of our
analysis. We then present in Section \ref{sect2:param} the results of
the atmospheric and physical parameters of all DB white dwarfs in our
sample, while objects of particular astrophysical interest are
discussed in Section \ref{sect2:Objects}. Finally, in the light of our
results, we discuss in detail the spectral evolution of DB white
dwarfs in Section \ref{sect2:SpectralEvolution}. Some concluding
remarks follow in Section \ref{sect2:conclusion}.

\section{DB White Dwarf Sample}\label{sect2:sample}

The first step of our investigation is to perform a full model
atmosphere analysis of DB white dwarfs using the photometric and
spectroscopic data from the SDSS database, as well as the
trigonometric parallaxes from the \textit{Gaia} survey. We describe
these spectroscopic, photometric, and astrometric samples in turn.

\subsection{Spectroscopic Sample}\label{sect2:specsample}

We retrieved the observed spectra of all the spectroscopically
identified DB white dwarfs in the SDSS, up to the DR12
\citep{DR7,DR10,DR12}. Since we want to characterize the entire
population of DB white dwarfs, we kept every subtype in our sample and
did not apply any criterion on the S/N value, but a visual inspection
of the spectroscopic fits allowed us to removed any problematic
data. We did however remove all objects with a spectral type
indicating a companion (M or +). Our final spectroscopic sample is
composed of 2058 spectra, representing 1915 individual white dwarfs,
since 128 of these have multiple spectroscopic observations. Except in
Section \ref{sect2:error_spec}, where each spectrum will be treated as
an independent object, we will retain only the best $\sn$ spectrum of
each object for our model atmosphere analysis.

\begin{figure}[t]
\centering
\includegraphics[clip=true,trim=3cm 9.5cm 3cm 4cm,width=0.8\columnwidth]{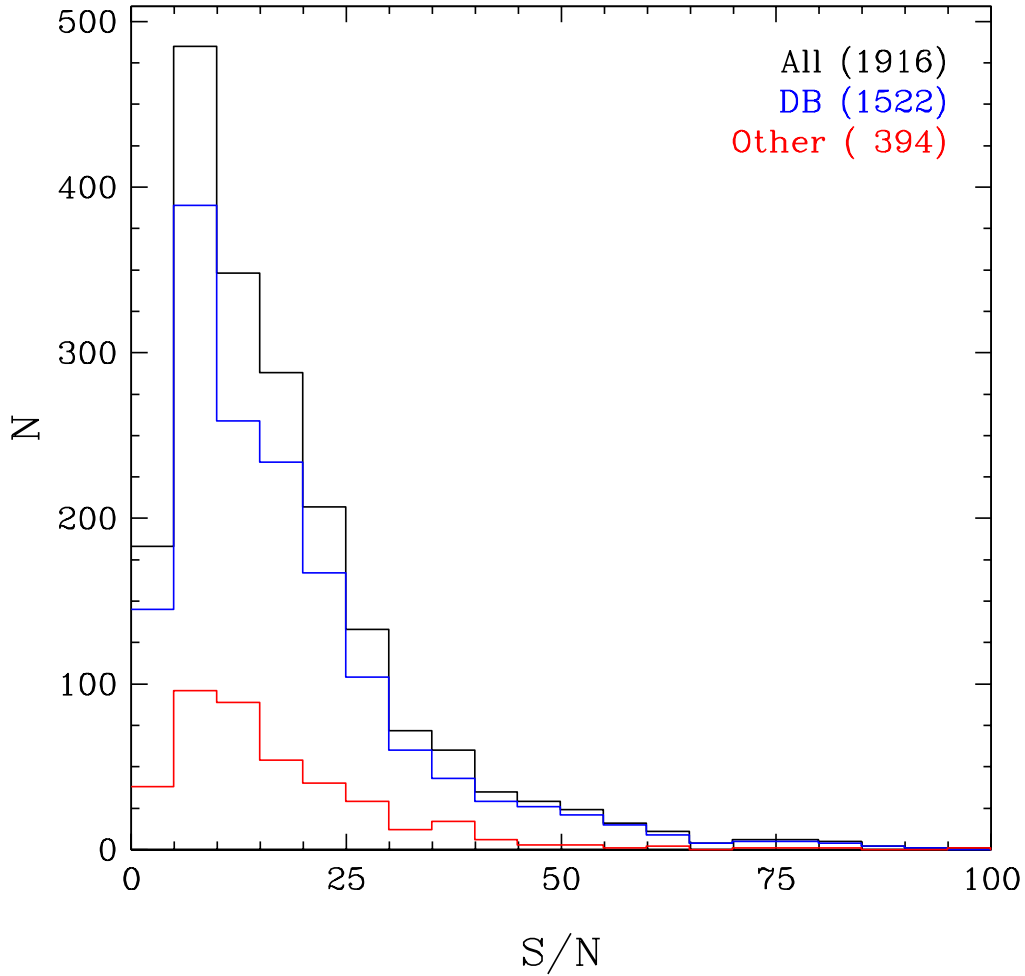}
\caption{Distribution of S/N of the complete spectroscopic sample (black), the DB subsample (blue), including the DB white dwarfs showing traces of hydrogen (DBA) and/or metals (DBZ), and other subtypes (red).}
\label{fig2:SN}
\end{figure}

Of the 1915 individual white dwarfs, 1522 (or 79.4\%) are classified
as DB, including the DB stars showing traces of hydrogen (DBA) and/or
metals (DBZ), while the other 20.6\% (394 objects) is composed of all
other subtypes, including magnetic objects (H), spectra with carbon
features (Q), and uncertain spectral types (:). Even though our
spectroscopic solution for these other subtypes might be uncertain, it
should not impact our conclusion significantly since they represent
only a small fraction of the entire sample. The S/N distribution of
our spectroscopic sample is presented in Figure \ref{fig2:SN}.

\subsection{Photometric and Astrometric Sample}\label{sect2:photsample}

Since an independent determination of the physical parameters can be
obtained from photometry, we also retrieved the $ugriz$ magnitudes for
all objects in our spectroscopic sample. We rely on the $ugriz$
photometric data from the SDSS 14$^{\rm th}$ data release (DR14),
since the calibration algorithm has been improved between DR7 and DR8,
and the $ugriz$ zero-points have been recalibrated in
DR13\footnote{https://www.sdss.org/dr14/algorithms/fluxcal/} as
well. We also want to take advantage of the $\gaia$ DR2 catalog
\citep{gaia2}, which provides precise trigonometric parallax
measurements for around 260,000 high confidence white dwarf candidates
\citep{Gentile2019}, and we thus retrieved the parallaxes for all DB
stars in our SDSS sample, if available (about 90\% of the objects in
our spectroscopic sample). Again here, we applied no specific
selection criteria on the quality of the $ugriz$ photometry or on the
trigonometric parallaxes, but we visually inspected all photometric
fits, and removed any obvious bad photometric or parallax data. As
before, we also removed all spectral types containing an M or a +. Our
final photometric sample is composed of 1669 photometric data sets, of
which 1350 (or $80.9\%$) are DB stars, including spectral types
indicating the presence of hydrogen (DBA) and/or metals (DBZ); the
other 19.1\% (319 photometric sets) is composed of all other subtypes
already mentioned above.

\begin{figure}[t]
\centering
\includegraphics[clip=true, trim=3cm 9.5cm 3cm 4cm,width=0.8\linewidth]{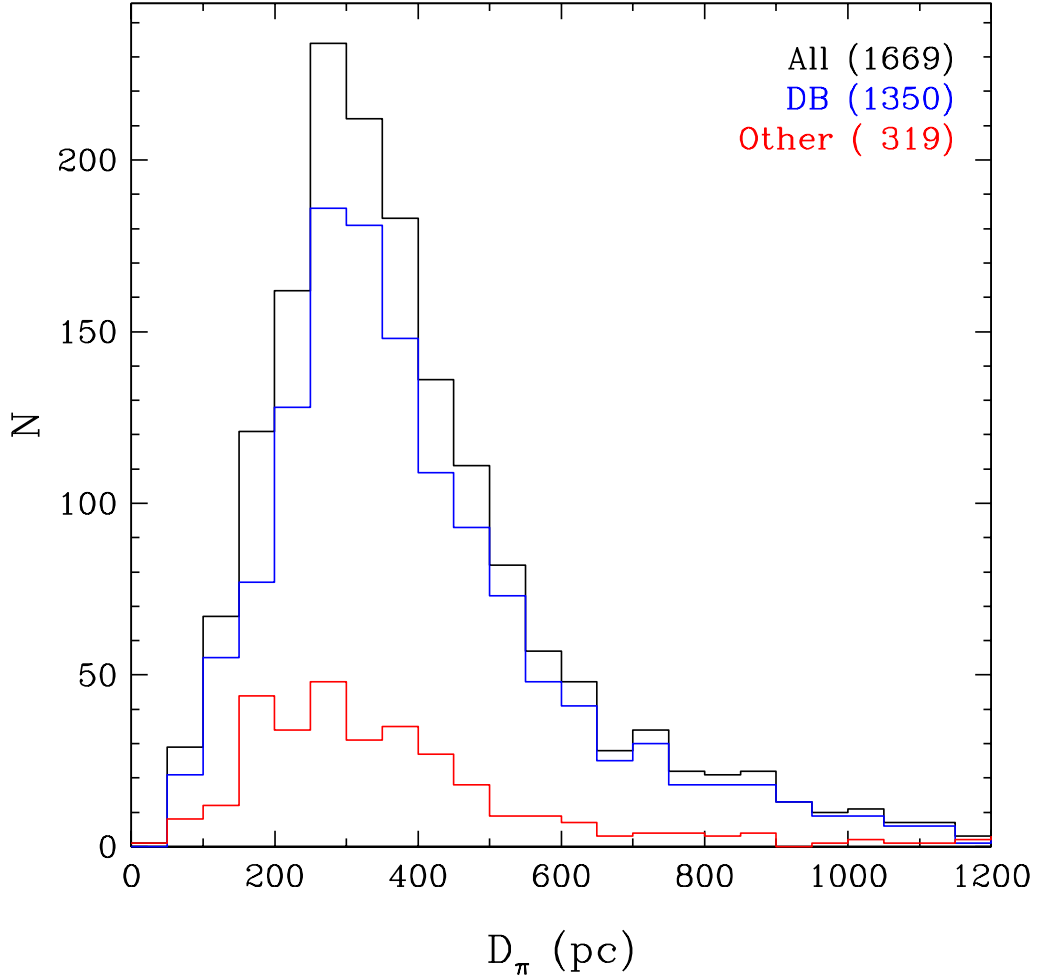}
\caption{Distribution of parallactic distances for the complete photometric sample (black), the DB and DBA subsample (blue), and the other subtypes (red).}
\label{fig2:distance}
\end{figure}

The distribution of white dwarfs in our sample as a function of
distance and spectral type is presented in Figure
\ref{fig2:distance}. As can be seen, most objects are located at very
large distances ($D > 100$~pc), which implies that their observed
$ugriz$ magnitudes will be significantly affected by interstellar
reddening. Interstellar extinction will be treated here (see also
GBB19 and \citealt{Bergeron2019}) following the procedure
outlined in \citet{Harris2006}, where the extinction is considered
negligible if $D \leq 100$ pc, to be maximum for the objects located
at $|z|> 250$ pc from the galactic plane, and to vary linearly between
these two regimes.

\section{Theoretical Framework}\label{sect2:theory}

The grid of model atmospheres used in the following photometric and
spectroscopic analyses is similar to that described in
\citet{Bergeron2011}, except for the treatment of van der Waals
broadening. We use here a treatment based on \citet{Deridder1976}
instead of \citet{Unsold1955}; see Sections 3.2 and 5.2 of GBB19 for
details. These models are in LTE and convection is treated with the
ML2/$\alpha$=1.25 version of the mixing-length theory (MLT). Our grid
covers a range of effective temperatures from 11,000 K to 50,000 K,
surface gravities from $\logg=7.0$ to $9.0$, and hydrogen abundances
(in number) from $\logH=-6.5$ to $-2.0$, as well as a pure helium
grid. Additional models have also been calculated that include the
Ca~\textsc{ii} H and K doublet, in order to properly fit the white
dwarfs in our sample showing strong calcium lines. This smaller grid
is similar to that described in \citet{Bergeron2011} --- except again
for the treatment of van der Waals broadening --- and covers a range
of $\Te=12,000~\K$ to 19,000 K, calcium abundances of $\logCa=-7.5$,
$-7.0$, $-6.5$, and $-6.0$, and the same range of $\logg$ and $\nHH$
values as before.

To determine the atmospheric and physical parameters of the DB white
dwarfs in our sample, we rely on the photometric and spectroscopic
techniques described at length in GBB19 and references
therein. Briefly, with the photometric approach, the effective
temperature $\Te$ and solid angle $\pi(R/D)^2$ are obtained by
comparing the observed energy distribution --- built from the $ugriz$
photometry --- with the predictions of model atmospheres, while with
the spectroscopic method, $\Te$, the surface gravity $\logg$, and the
hydrogen abundance $\nHH$ are obtained by comparing the observed and
synthetic spectra, both normalized to a continuum set to
unity. Stellar masses can then be derived from evolutionary models. We
rely here on C/O-core envelope models\footnote{See
  http://www.astro.umontreal.ca/$\sim$bergeron/CoolingModels.} similar
to those described in \citet{Fontaine2001} with thin hydrogen layers
of $q({\rm H})\equiv M_{\rm H}/M_\star=10^{-10}$, which are
representative of helium-atmosphere white dwarfs. For the DB stars in
our sample showing calcium lines, we explore the spectroscopic fits
obtained for each calcium abundance in our grid, and adopt the
solution with the lowest $\chi^2$ value.

\section{Error Estimation}\label{sect2:error}

As discussed in the Introduction, our understanding of the nature and
evolution of DB white dwarfs rests heavily on our ability to measure
their atmospheric and physical parameters with great accuracy and
precision. We thus present in this section a thorough analysis of the
photometric and spectroscopic errors associated with the
determinations of white dwarf parameters.

\subsection{Photometric Errors}\label{sect2:error_phot}

The errors associated with our photometric solutions can be obtained
directly from the covariance matrix of the Levenberg-Marquadt
minimization procedure used in the photometric technique. These depend
mostly on the uncertainty associated with the observed magnitudes and
the sensitivity of the $ugriz$ photometry to the atmospheric
parameters. We apply here a lower limit of 0.03 mag on each bandpass
so that the photometric solution is not driven by a single magnitude
with an extremely small uncertainty (see also
\citealt{Bergeron2019}). As also discussed in \citet{Bergeron2019},
the SDSS magnitude system is not exactly on the AB magnitude system,
and corrections of the order of 0.03 mag must be added to some
bandpasses, all of which remain uncertain. Since we use the
trigonometric parallax in the fitting procedure, the errors on the
atmospheric parameters will depend on $\sigma_\pi$ as well. Note that
all values reported in this section are for the DB stars only, and we
expect our errors for other subtypes (e.g. magnetic DB stars or
uncertain spectral types) to be even larger.

\begin{figure*}[t]
\centering
  \includegraphics[clip=true,angle=270,trim=2cm 1.5cm 2cm 2cm,width=\linewidth]{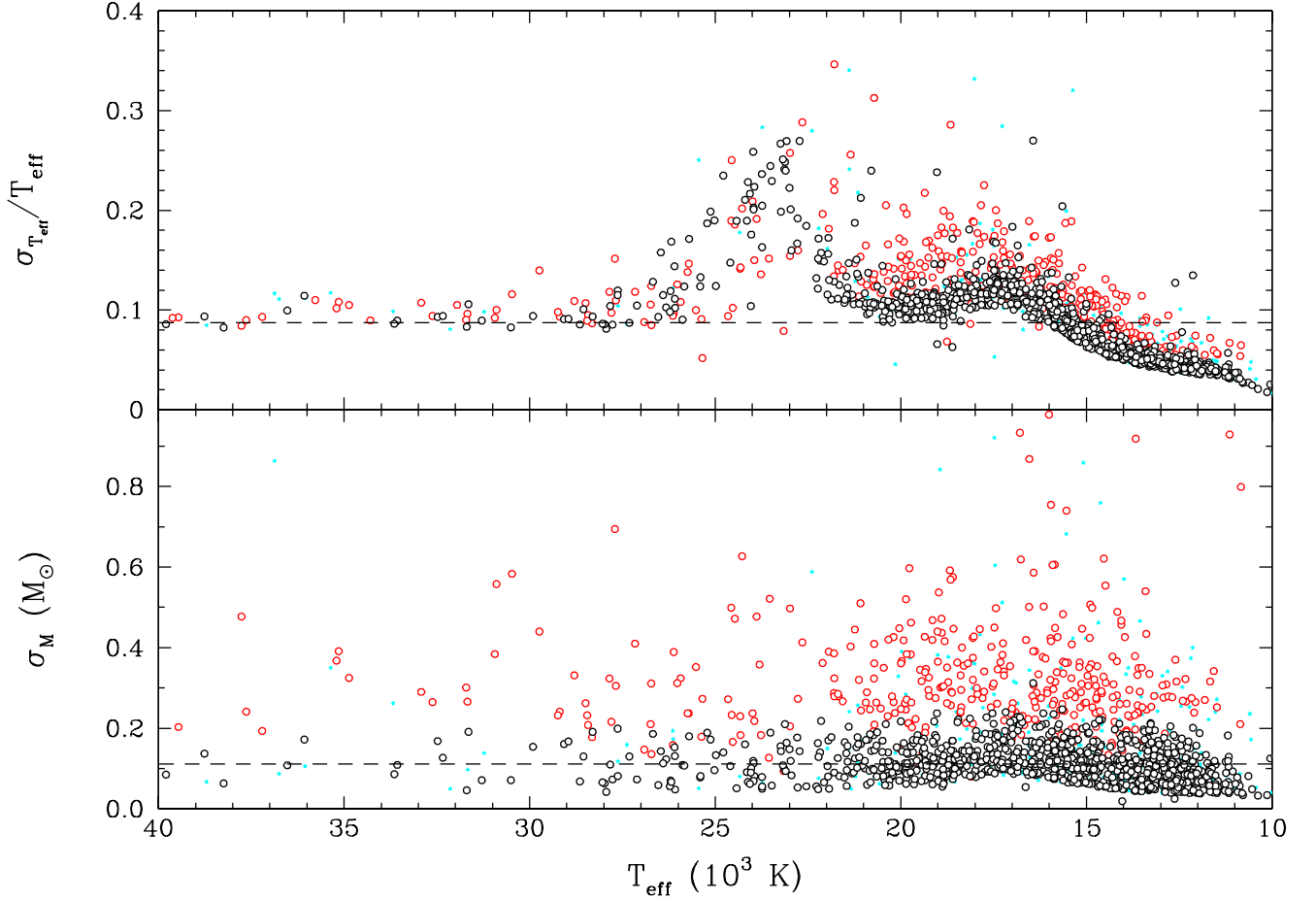}
  \caption{Distribution of errors on $\Te$ (top panel) and stellar
    mass (bottom panel) obtained from the photometric technique, as a
    function of effective temperature. The open circles represent the
    DB and DBA stars in our sample with $\sigma_\pi/\pi < 0.25$
    (black) and $\sigma_\pi/\pi>0.25$ (red). The dashed lines indicate
    the mean errors of the DB/DBA photometric subsample with
    $\sigma_\pi/\pi < 0.25$. All other spectral types are represented
    by cyan dots.}
  \label{fig2:phot_errors}
\end{figure*}

\begin{figure*}[!t]
    \centering
    \includegraphics[clip=true,trim=2.1cm 3.5cm 2.5cm 2.5cm,width=0.8\linewidth]{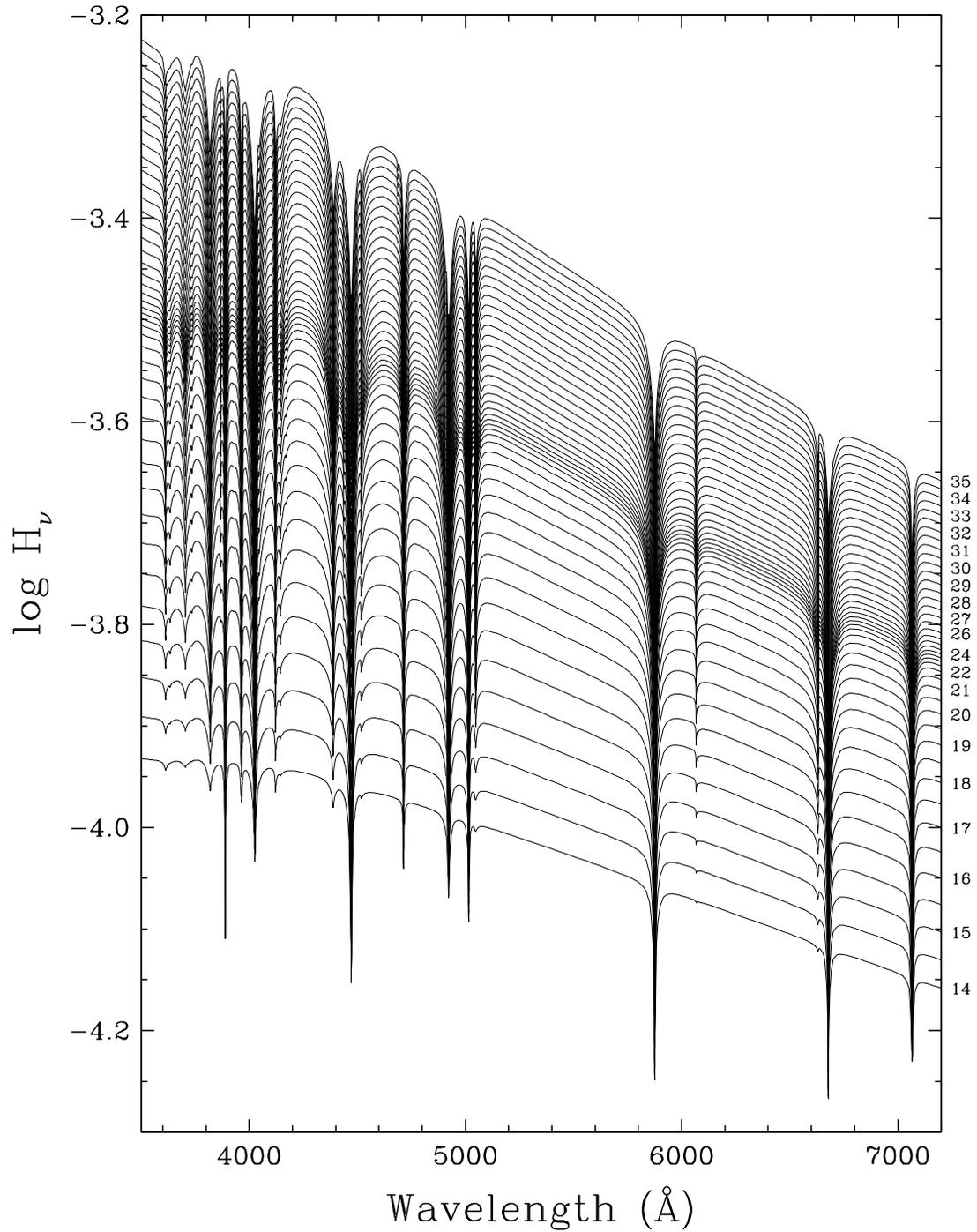}
    \caption{Eddington fluxes (in units of ergs cm$^{-2}$ s$^{-1}$
      Hz$^{-1}$) as a function of wavelength for pure helium DB models
      at $\logg=8$, and for various effective temperatures (in units
      of $10^{3}$~K) indicated in the figure.}
    \label{fig2:plotsynthlog}
\end{figure*}

The errors associated with our photometric effective temperatures are
presented in the top panel of Figure \ref{fig2:phot_errors}. The mean
error for the overall sample is $\sim$10\% in $\Te$, but the
individual errors vary significantly as a function of temperature
since they widely depend on the sensitivity of the $ugriz$ photometry
to variations in $\Te$ (see Figure 4 of GBB19). For instance, the
errors drop significantly below $\sim$16,000~K, where the $ugriz$
photometry is very sensitive to $\Te$, reaching values as small as 2\%
near 10,000 K. More puzzling is the significant increase in
$\sigma_{\Te}$ in the range $26,000~\K>\Te>22,000~\K$. We traced back
this feature --- never discussed in the literature, to our knowledge
--- to the particular behavior of the Eddington fluxes in the optical
regions, which increase very slowly in this temperature
range\footnote{The range of temperature at which this behavior is
  observed is actually a function of $\logg$.}, when the main opacity
source switches from the He~\textsc{i} bound-free opacity to the
He~\textsc{ii} free-free opacity, as illustrated in Figure
\ref{fig2:plotsynthlog}. Finally, $\sigma_{\Te}$ does not appear to be
significantly affected by the parallax uncertainties $\sigma_\pi$. If
we restrict our sample to the objects for which $\sigma_\pi/\pi<0.25$,
we obtain $\langle \sigma_{\Te} \rangle = 8.7\%$, a value only
slightly lower than that obtained above. This is an expected result
since $\Te$ is determined mainly from the shape of the energy
distribution, which does not depend on the parallactic distance (after
dereddening).

The errors associated with photometric masses are presented in the
bottom panel of Figure \ref{fig2:phot_errors}. The mean error for the
overall sample is somewhat high, $\langle \sigma_M \rangle= 0.341$
$\msun$, but this is mostly caused by the objects in our sample with
very large parallax uncertainties. Unlike for the effective
temperature, the precision on the mass relies heavily on $\sigma_\pi$
since the parallactic distance is used to obtain the radius from the
solid angle, which is then converted into mass using the mass-radius
relation (see section 4.1 of GBB19). If we restrict our sample to
objects with $\sigma_\pi/\pi<0.25$, we obtain a much lower mean value
for the error of $\langle \sigma_M \rangle=0.112~\msun$. We also note
that the $\sigma_M$ distribution in Figure \ref{fig2:phot_errors} is
fairly constant with effective temperature, unlike what is observed
for $\sigma_{\Te}$.

Finally, if we do not apply the 0.03 mag lower limit uncertainty on
the $ugriz$ magnitudes, the mean uncertainties are only slightly
lower: $\langle \sigma_{\Te} \rangle =6.85\%$ and $\langle \sigma_{M}
\rangle =0.102~\msun$ (for $\sigma_\pi/\pi<0.25$).

\subsection{Spectroscopic Errors}\label{sect2:error_spec}

\begin{figure*}[t]
\centering
  \includegraphics[clip=true,angle=270,trim=2cm 2cm 1.5cm 2cm,width=\linewidth]{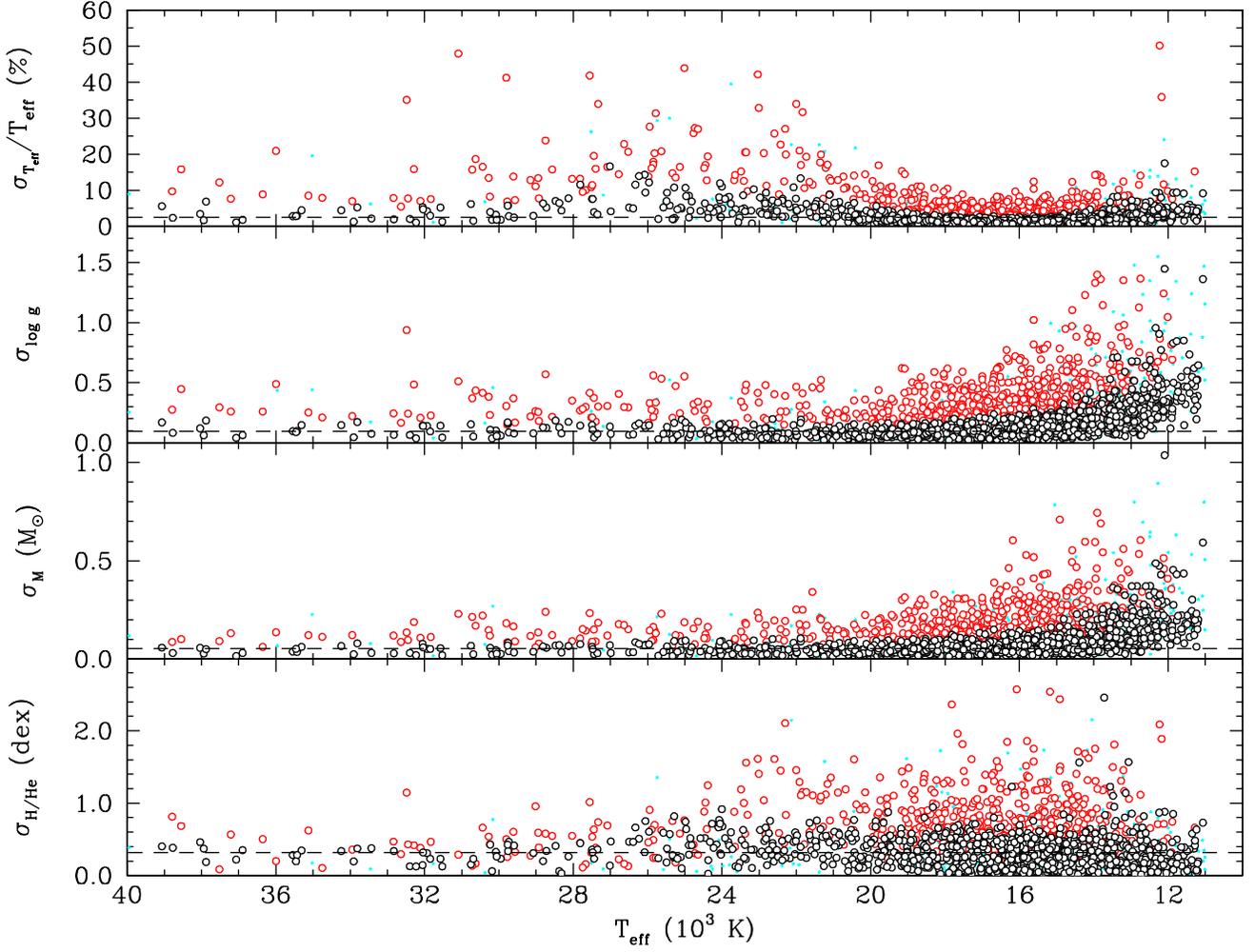}
  \caption{Internal errors on $\Te$, $\logg$, $M$, and $\nHH$ obtained
    from the spectroscopic technique, as a function of effective
    temperature. The open circles represent the DB(Z) and DBA(Z) white
    dwarfs with $\sn>10$ (black) and $\sn<10$ (red); all other
    subtypes are represented by cyan dots. The dashed lines indicate
    the corresponding mean errors for the objects with $\sn>10$, and
    $\Te>16,000$~K for $\sigma_{\logg}$ and $\sigma_{M}$.}
  \label{fig2:int_errors}
\end{figure*}

The spectroscopic errors associated with the atmospheric parameters
--- $\Te$, $\logg$, and $\nHH$ --- can be estimated from the
covariance matrix of the Levenberg-Marquadt procedure used in our
fitting technique. These so-called {\it internal errors} represent the
ability of our models to reproduce the data, and can be made
arbitrarily small if the $\sn$ of the spectra is high enough (see
\citealt{LBH05} for a full discussion). The internal errors on $\Te$,
$\logg$, and $\nHH$ are displayed in Figure \ref{fig2:int_errors} as a
function of effective temperature, together with the errors on the
spectroscopic mass, obtained by combining the effects of
$\sigma_{\Te}$ and $\sigma_{\logg}$ on the mass determination (note
that $\sigma_M$ is completely dominated by the contribution of
$\sigma_{\logg}$, however).

The mean error on $\Te$ for the overall sample is $\langle
\sigma_{\Te}\rangle =4.43\%$, or 2.59\% if we restrict our sample to
$\sn>10$. The individual errors $\sigma_{\Te}$ (in percentage) are
fairly constant with effective temperature, except between
$\sim$20,000 K and 30,000 K where the helium lines become less
sensitive to $\Te$ (see Figure 1 of \citealt{Bergeron2011}). The mean
error on $\log g$ is $\langle \sigma_{\logg} \rangle = 0.263$, and on
the mass $\langle \sigma_M \rangle = 0.156~\msun$; these values drop
to 0.163 and $0.088~\msun$, respectively, if we restrict our sample to
$\sn>10$. Unlike for $\sigma_{\Te}$, there is no increased scatter in
the distributions between 20,000 K and 30,000 K, indicating that the
helium lines remain sensitive to $\log g$ and mass in this temperature
range. However, we can see that the individual errors become more
important below $\Te\sim16,000~\K$, where neutral broadening remains a
large source of uncertainty in our models (see GBB19 and references
therein). If we exclude the objects below $16,000~\K$, the mean errors
drop even further to $\langle \sigma_{\logg} \rangle = 0.094$ and
$\langle \sigma_M \rangle = 0.053~\msun$ (for $\sn>10$). Finally, the
mean error on the hydrogen abundance is $\langle \sigma_{\rm H/He}
\rangle = 0.486$ dex (or 0.314 dex for the sample with $\sn>10$), but
this somewhat large mean value is dominated by the objects for which
we could only determine an upper limit on the hydrogen abundance. If
we restrict our estimation to the DBA stars in our sample, the mean
error drop to 0.298 dex (or 0.213 dex for the sample with $\sn>10$).

\begin{figure*}[t]
\centering
  \includegraphics[clip=true,trim=1cm 8.5cm 1cm 1cm,width=0.92\linewidth]{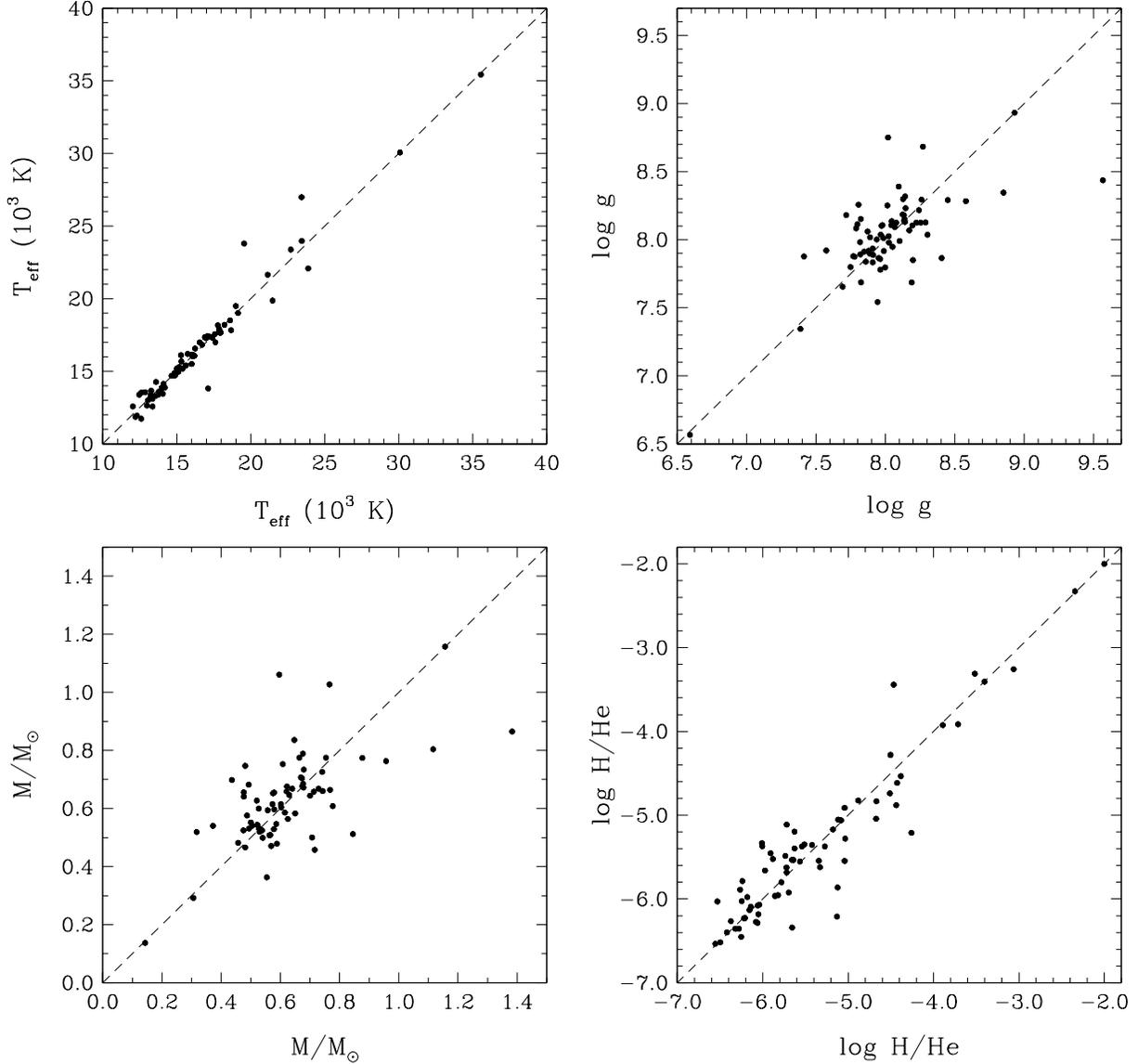}
  \caption{Comparison of $\Te$, $\logg$, $M$, and $\logH$ for the 49 objects in our sample with multiple spectroscopic observations and with $\sn>10$.}
  \label{fig2:ext_errors}
\end{figure*}

Another way to estimate the errors using the spectroscopic technique
is from multiple observations of the same star, as described for
instance in \citet{LBH05} and \citet{Bergeron2011}. To get a good
estimate of the external errors, we excluded all spectra with $\sn<10$
as well as uncertain spectral types, which left us with 49 objects
with multiple spectra. For those with more than two observations, we
kept only the two highest $\sn$ spectra. These so-called
\textit{external errors} are displayed in Figure
\ref{fig2:ext_errors}. The mean external errors are $\langle\Delta
\Te/\Te\rangle=2.40\%$, $\langle \Delta\logg\rangle = 0.152$, $\langle
\Delta M \rangle= 0.086$ $\msun$, and $\langle \Delta {\rm H/He}
\rangle =0.199$ dex. These values compare favorably well with the
internal errors discussed above. \citet{Bergeron2011} obtained smaller
values of $\langle \Delta \Te/\Te \rangle = 2.3\%$ and $\langle \logg
\rangle = 0.052$, which can be explained by the fact that the spectra
used in their analysis had $\sn > 50$, while in our restricted sample
$\langle \sn \rangle \sim 25$ (see Figure \ref{fig2:SN}). Perhaps a
more useful comparison is with the values obtained by
\citet{Koester2015} based on multiple spectra from the SDSS. They
obtained similar mean errors of 3.1\% in $\Te$, 0.12 in $\logg$, and
0.18 dex in $\nHH$.

We reevaluate again in Section \ref{sect2:accuprec} the precision and
accuracy of the photometric and spectroscopic techniques, but only
after we compare the atmospheric and physical parameters determined
using both fitting methods.

\section{Atmospheric and Physical Parameters of DB White Dwarfs}\label{sect2:param}

Our main goal is to characterize the entire DB white dwarf population
in the SDSS. To do so, we take advantage of the photometric and
spectroscopic techniques to obtain independent determinations of the
atmospheric and physical parameters of each star, such as the
effective temperature $\Te$, the surface gravity $\logg$, the stellar
mass $M$, and the hydrogen abundance ratio $\nHH$. Note that the
latter can only be determined spectroscopically. The derived
parameters are provided in Table \ref{tab1}. Even though
our sample contains several subtypes, particular attention will be
given to the DB and DBA white dwarfs (with or without metals) as these
represent about 80\% of our sample (see Section
\ref{sect2:sample}). In this section, we discuss in turn the surface
gravity, stellar mass, and photospheric hydrogen abundance
distributions.

\begin{deluxetable*}{c|cccc|cccrcc|c}[t]
\tabletypesize{\scriptsize}

\tablecaption{List of photometric/spectroscopic atmospheric/physical parameters \label{tab1}}

\tablehead{\colhead{} & \multicolumn{4}{c}{Photometry} & \multicolumn{6}{c}{Spectroscopy} & \colhead{}\\
\colhead{SDSS name} & \colhead{$\Te$} & \colhead{$\logg$} & \colhead{$M$} & \colhead{$\sigma_\pi/\pi$} & \colhead{$\Te$} & \colhead{$\logg$} & \colhead{$M$} & \multicolumn{1}{c}{$\logH$} & \colhead{$\logCa$} & \colhead{$\sn$} & \colhead{Notes} \\
      \colhead{} & \colhead{(K)}    & \colhead{}  & \colhead{($\msun$)} &\colhead{} & \colhead{(K)} & \colhead{} & \colhead{($\msun$)} & \colhead{} & \colhead{} & \colhead{}& \colhead{}}

\startdata 
$000055.12-042449.00$ &14,675 &  7.80 & 0.483 &  0.694 & 13,649 &  7.88 & 0.517 & $ -5.111$ & \nodata &   10.16&\\            
$000106.22+250330.00$ &15,753 &  7.03 & 0.237 &  0.527 & 14,836 &  7.88 & 0.520 & $ -4.748$ &   -7.50 &   18.68&\\        
$000111.66+000342.55$ & \nodata  & \nodata   & \nodata   & \nodata    & 11,205 &  6.32 & 0.022 & $ -5.327$ & \nodata &   18.84& 1 \\         
$000116.49+000204.45$ &11,238 &  7.87 & 0.510 &  0.093 & 11,304 &  7.73 & 0.440 & $ -5.965$ & \nodata &   31.52&1\\           
$000122.51+235934.20$ & \nodata  & \nodata   & \nodata   & \nodata    & 12,339 &  6.88 & 0.194 & $< -5.672$ & \nodata &    7.63& 1 \\        
$000205.57+002041.80$ &19,926 &  7.11 & 0.266 &  1.388 & 18,108 &  8.16 & 0.693 & $ -5.025$ & \nodata &    6.47&\\            
$000223.06+272358.50$ &16,437 &  8.08 & 0.637 &  0.033 & 16,830 &  8.00 & 0.594 & $< -6.133$ & \nodata &   49.67&\\           
$000407.15+264939.70$ &12,800 &  7.96 & 0.565 &  0.222 & 14,050 &  9.28 & 1.288 & $ -5.263$ & \nodata &    9.47&\\            
$000426.95+243258.90$ &10,564 &  8.25 & 0.738 &  0.173 & 16,206 &  9.16 & 1.253 & $ -5.090$ &   -6.00 &   10.46&\\        
$000447.06+240703.60$ & \nodata  & \nodata   & \nodata   & \nodata    & 15,328 &  7.75 & 0.457 & $ -5.434$ & \nodata &    9.47& \\           
$000509.94+003809.60$ &47,176 &  7.68 & 0.503 &  0.278 & 50,033 &  7.68 & 0.509 & $< -3.280$ & \nodata &   16.24&\\           
$000515.58+071313.71$ &17,994 &  7.90 & 0.538 &  0.057 & 19,000 &  8.15 & 0.686 & $< -5.600$ & \nodata &   40.62&\\           
$000720.22-002325.40$ &19,434 &  8.10 & 0.658 &  0.096 & 19,099 &  7.90 & 0.540 & $< -5.383$ & \nodata &   18.67&\\           
$000730.75+275111.90$ &13,932 &  7.64 & 0.405 &  0.250 & 16,305 &  8.06 & 0.629 & $ -5.880$ & \nodata &   11.92&2\\           
$000731.17-095849.10$ &18,651 &  7.90 & 0.541 &  0.326 & 20,170 &  8.05 & 0.625 & $< -4.849$ & \nodata &    6.32&\\           
$000742.62+252422.50$ &20,913 &  8.64 & 0.999 &  0.368 & 28,723 &  7.80 & 0.511 & $< -3.702$ & \nodata &   12.85&\\           
\enddata
\tablenotetext{1}{Marginal helium lines}
\tablenotetext{2}{DB+DB unresolved double degenerate candidate}
\tablenotetext{3}{DA+DB unresolved double degenerate system}
\tablenotetext{4}{Objects excluded from the analysis}
\tablecomments{Table \ref{tab1} is published in its entirety in the machine-readable format. 
A portion is shown here for guidance regarding its form and content.}

\end{deluxetable*}

\subsection{Surface Gravity Distributions}\label{sect2:logg}

\begin{figure*}[t]
\centering
  \includegraphics[clip=true,angle=270,trim=2cm 3cm 2.5cm 2cm,width=0.95\linewidth]{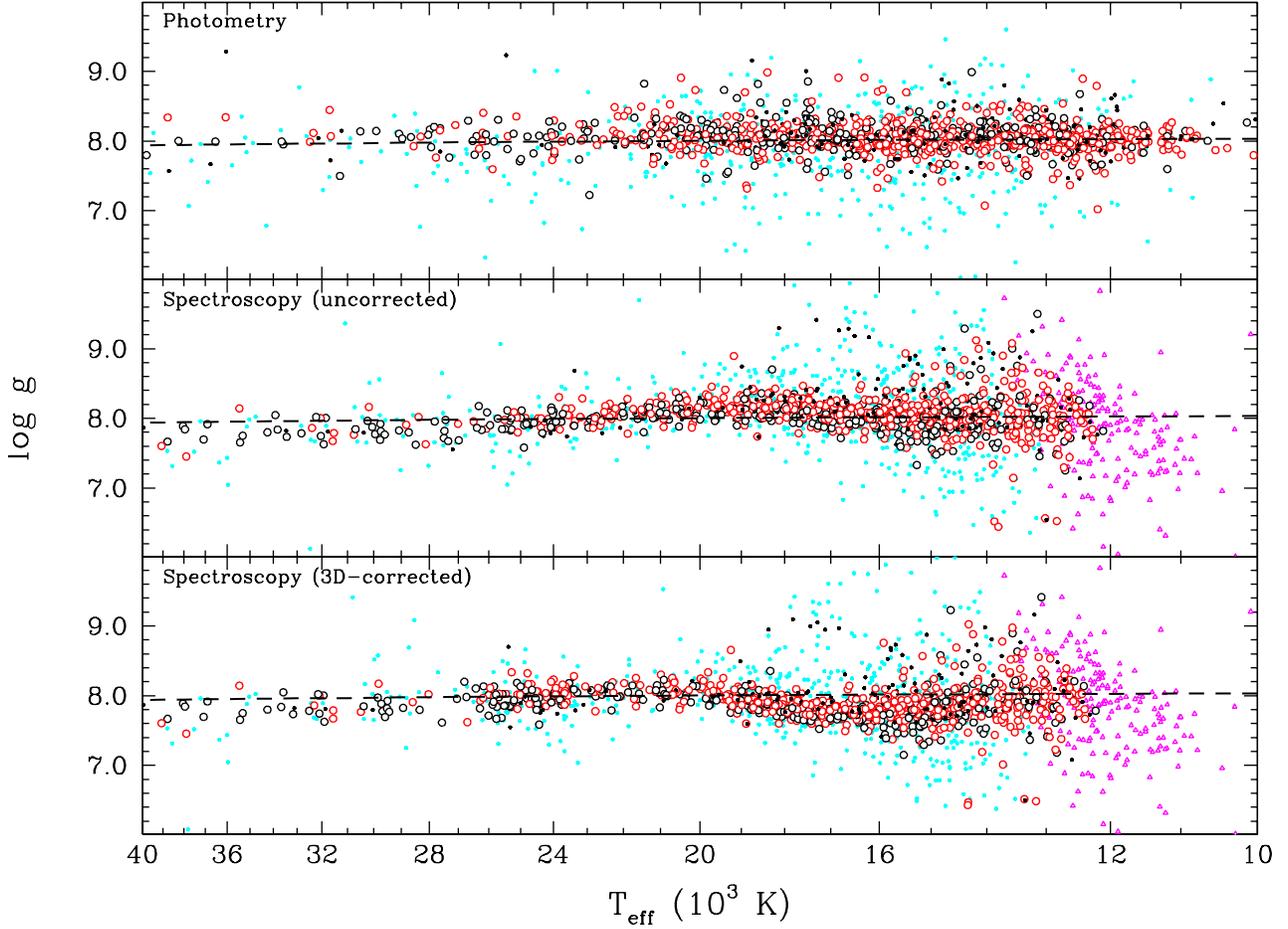}
  \caption{Photometric (upper panel) and spectroscopic (middle panel)
    $\logg$ distributions as a function of $\Te$. The open circles
    represent the DB (black) and DBA (red), with or without metals,
    while the black dots correspond to other spectral types with
    $\sigma_\pi/\pi<0.25$ (photometry) or $\sn>10$
    (spectroscopy). Spectra with marginal helium lines are shown as
    magenta triangles, and any object with $\sigma_\pi/\pi>0.25$ or
    $\sn<10$, regardless of the spectral type, is represented by a
    cyan dot. Bottom panel: Same as middle panel, but with the 3D
    hydrodynamical corrections taken into account. In all panels, the
    dashed line represents a 0.6 $\msun$ evolutionary sequence.}
  \label{fig2:logg3D}
\end{figure*}

We present in Figure \ref{fig2:logg3D} the photometric and
spectroscopic surface gravities as a function of effective temperature
for all DB white dwarfs in our sample. The photometric $\logg$
distribution (upper panel) is well centered on the 0.6 $\msun$
evolutionary sequence at all temperatures, as expected. Moreover, the
dispersion in $\logg$ values appears fairly constant with $\Te$, and
the distribution in temperature is also uniform, in the sense that
there are no regions with an obvious accumulation or depletion of
objects. The spectroscopic $\logg$ distribution (middle panel) is also
uniform in temperature, but contrary to the photometric distribution,
it deviates significantly from the 0.6 $\msun$ sequence in several
temperature ranges. First, at the very cool end of the distribution
($\Te\lesssim13,000$ K), the $\logg$ values are much lower than the
canonical $\log g=8$ value. This behavior can be partially explained
by the weakness of the helium lines in this temperature range (see
Figure 3 of \citealt{Rolland2018}), and the spectroscopic technique
has most likely reached its limits below which the atmospheric
parameters become unreliable. We estimated this limit at the
temperature where the equivalent width of the He~\textsc{i}
$\lambda4471$ line becomes smaller than 3 \AA~(4 \AA, 5 \AA) for
$\sn>20$ ($10<\sn<20$, $\sn<10$). These objects are shown as magenta
triangles in Figure \ref{fig2:logg3D} and will not be considered any
further in our analysis.

\begin{figure*}[t]
\centering
  \includegraphics[angle=270,clip=true,trim=2cm 2cm 3cm 2cm,width=\linewidth]{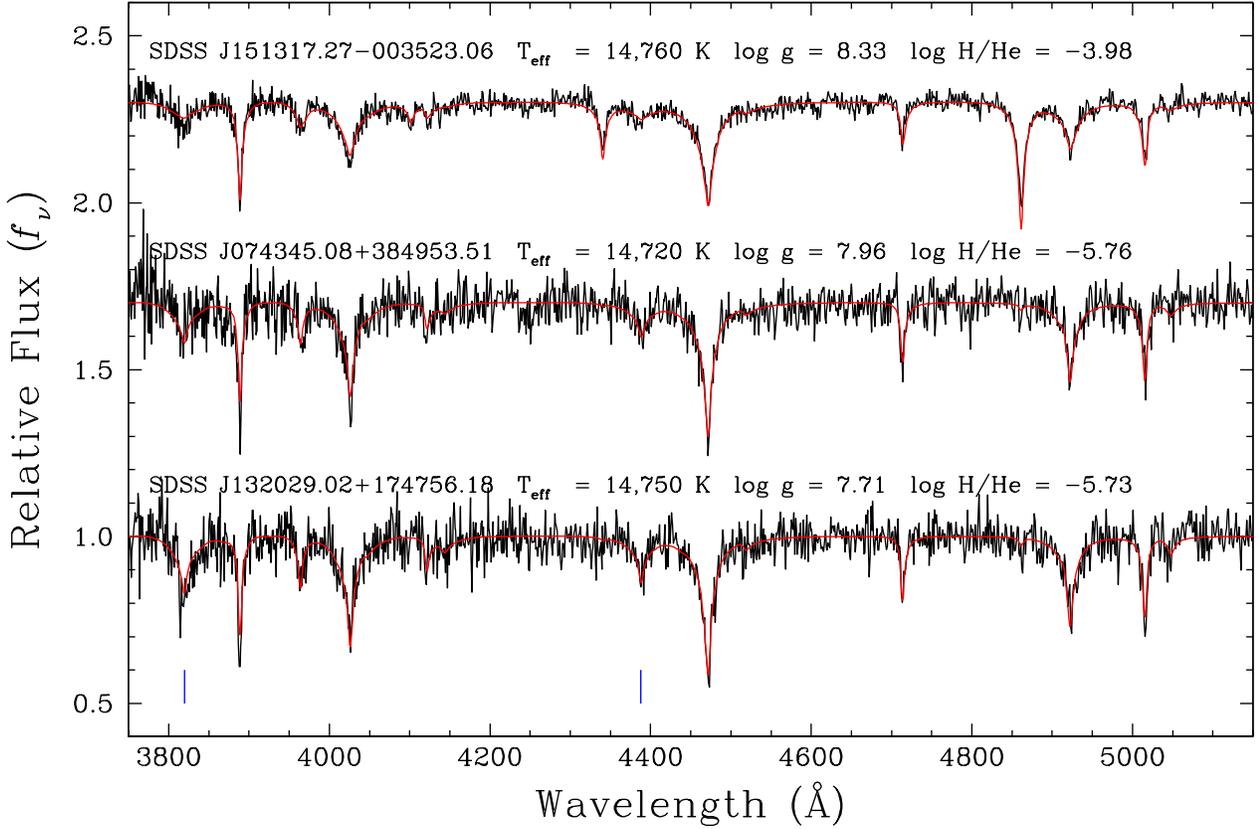}
  \caption{Our best spectroscopic fits to 3 DB(A) white dwarfs near
    $\Te=14,700~\K$, but with significantly different surface
    gravities. The spectra have been normalized to a continuum set to
    unity and the best-fit solutions are shown by the red lines. The
    atmospheric parameters for each object are indicated as well. The
    location of He~\textsc{i} $\lambda$3820 and $\lambda$4388, which
    are the most $\logg$-sensitive in this temperature regime, is
    shown by the blue tick marks.}
  \label{fig2:difflogg}
\end{figure*}

However, even by removing from the sample these spectra with
marginally detectable helium lines, we still see a larger scatter
around $\Te\sim13,000-16,000$ K, in contrast to what is observed in
photometry. The main line broadening mechanism in this particular
temperature range is the broadening by neutral particles, or more
specifically van der Waals broadening. As mentioned in several
studies, the large scatter at the cool end of the spectroscopic $\log
g$ distribution ($\Te<16,000~\K$) is mostly caused by the improper
treatment of van der Waals broadening
(\citealt{Beauchamp1996,Bergeron2011,Koester2015}; GBB19), and the
theory currently used in our model atmospheres probably still requires
some improvement in order to determine more accurate $\logg$ values
below $\Te = 16,000~\K$. Nevertheless, we do find convincing cases of
white dwarfs in this temperature range with a broad range of
spectroscopic $\logg$ values. This is illustrated in Figure
\ref{fig2:difflogg}, where we show 3 DB(A) white dwarfs with similar
spectroscopic temperatures ($\Te \sim 14,700~\K$), but with different
surface gravities, ranging from $\log g = 7.7$ to 8.3 (see also Figure
7 of \citealt{Limoges2010} and the related discussion). For these 3
objects, the shape and strength of the He~\textsc{i} $\lambda$3820 and
$\lambda$4388 lines, which are particularly $\logg$-sensitive in this
temperature regime, vary significantly. We would like to stress that
this is independent of any theoretical modeling, since it is observed
directly in the spectrum. This suggests that, despite the
unsatisfactory treatment of van der Waals broadening, at least part of
the scatter in $\logg$ observed below $\Te = 16,000~\K$ might be real
after all. This is supported by the fact that the photometric
distribution also shows several high-$\logg$ objects. We come back to
this point further in Section \ref{sect2:mass}.

At the hot end ($\Te>27,000~\K$) of the spectroscopic $\logg$
distribution in Figure \ref{fig2:logg3D}, we see also a trend towards
lower $\logg$ values, with inferred masses below 0.6 $\msun$, again in
sharp contrast with what is observed in photometry\footnote{Note that
  these hot objects in spectroscopy appear at lower photometric
  temperatures when the energy distribution sampled by the $ugriz$
  photometry is in the Rayleigh-Jeans regime (see Figure 13 of GBB19),
  but this barely affects the photometric masses, as shown in Figure
  17 of GBB19 (see also Figure 7 of
  \citealt{Bergeron2019}).}. \citet{Tremblay2011} and
\citet{Genest2014} observed a similar phenomenon when analyzing the
spectroscopic $\logg$ distribution of DA white dwarfs from the
SDSS. In this case, a comparison with the DA spectra taken from
\citet[][see for instance Figures 14 and 15 of
  \citealt{Genest2014}]{Gianninas2011}, {\it where this effect was not
  observed}, indicated that the apparent decrease in $\logg$ when
using the SDSS spectra could be attributed to residual flux
calibration issues. Since this calibration problem must also affect
the spectra of DB white dwarfs, we conclude that the apparent decrease
in spectroscopic $\logg$ values observed in Figure \ref{fig2:logg3D}
above $\Te\sim27,000~\K$ is most likely an artifact caused by
calibration issues with the SDSS spectra.

Another possibility, mentioned by \citet{Koester2015}, is that the
cooler white dwarfs might originate from more massive progenitors,
since DB stars at $\Te\sim10,000~\K$ are about $5 \times 10^8$ years
older than those at $\Te\sim30,000~\K$. They estimated that the hotter
stars could be less massive than the cooler ones by about 0.05
$\msun$, thus explaining the slight decrease in $\logg$. However, even
though this would be a valid explanation in a stellar cluster where
there is a single burst of star formation, it certainly does not apply
in the case of field white dwarfs where continuous star formation
occurs. And indeed, such an effect in $\logg$ (or mass) is not
detected in the DA population (see Figure 30 of
\citealt{Gianninas2011} for instance).

\begin{figure*}[t]
\centering
  \includegraphics[clip=true,angle=270,trim=1cm 3cm 2.5cm 1cm,width=0.92\linewidth]{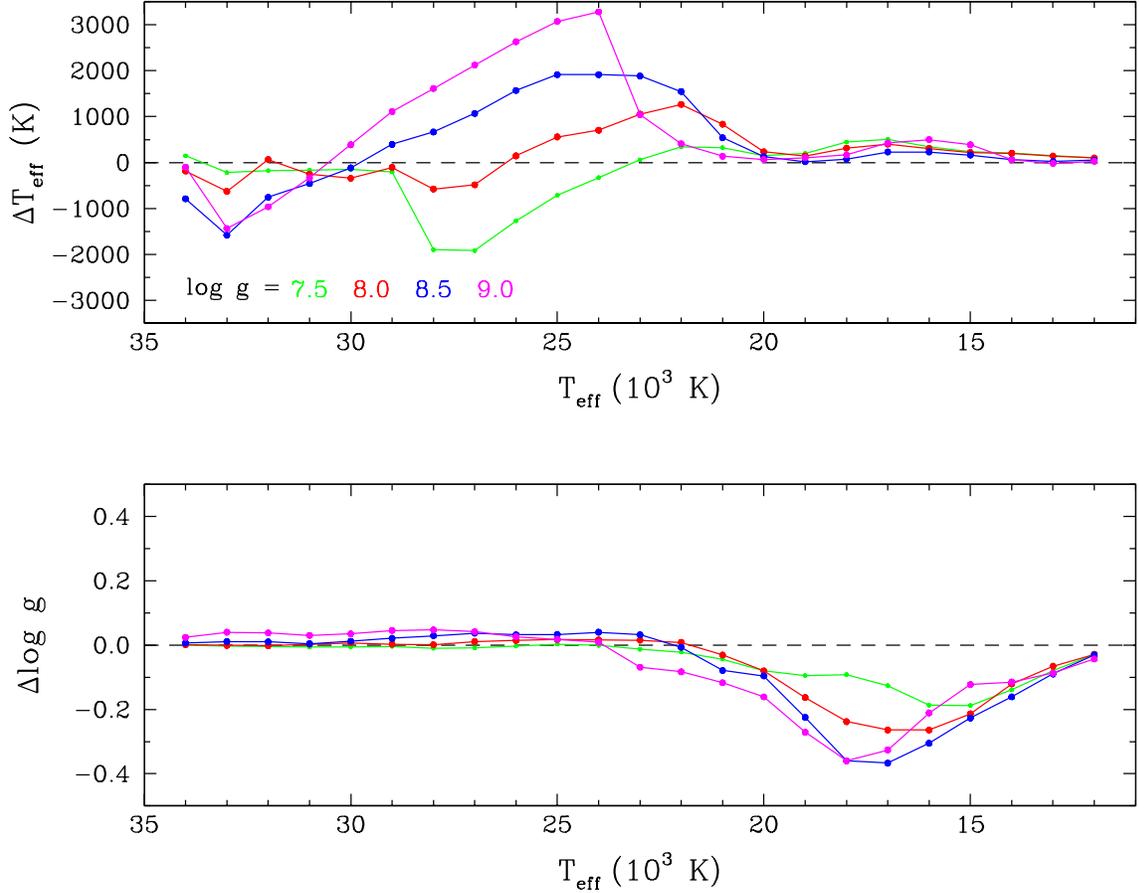}
  \caption{Theoretical 3D hydrodynamical corrections in $\Te$ (top
    panel) and $\logg$ (bottom panel) to be applied to 1D
    spectroscopic solutions, as a function of effective temperature
    and for various $\logg$ values, as described in Table 2 of
    \citet{Cukanovaite2018}.}
  \label{fig2:corr3D}
\end{figure*}

Finally, the spectroscopic $\logg$ distribution in Figure
\ref{fig2:logg3D} also shows a small but significant increase in the
range $22,000~\K>\Te>16,000~\K$, also noted by GBB19. In this region,
we obtain from spectroscopy $\langle \logg \rangle = 8.15$ (for
$\sn>10$), a value slightly larger than that obtained from photometry,
$\langle \logg \rangle = 8.08$ (for $\sigma_\pi/\pi < 0.25$). This
corresponds to the temperature range where the 3D hydrodynamical
effects become important according to \citet{Cukanovaite2018}, who
recently calculated 3D model atmospheres for pure helium-atmosphere
white dwarfs. Cukanovaite et al.~also published 3D corrections (see
their Table 2) to be applied to the 1D spectroscopic solutions
(assuming ML2/$\alpha=1.25$), which are reproduced here in Figure
\ref{fig2:corr3D}, for completeness. The largest corrections in
$\logg$ occur near 17,000 K --- which incidentally coincides with the
maximum increase in $\logg$ observed in the middle panel of Figure
\ref{fig2:logg3D} ---, while the largest corrections in $\Te$ occur at
much higher temperatures, near $\Te=25,000~\K$.

The 3D-corrected spectroscopic $\logg$ distribution for our sample is
displayed in the bottom panel of Figure \ref{fig2:logg3D}. While the
mean $\logg$ value between 20,000 K and 22,000 K is now 8.05 --- in
excellent agreement with the photometric mean --- the surface
gravities below this temperature range are over-corrected when
compared to the photometric results. For instance, between
$\Te\sim15,000~\K$ and 20,000~K, the spectroscopic distribution has a
mean value of $\logg=7.87$ (for $\sn>10$), which is 0.18 dex below the
corresponding photometric value, suggesting that the $\logg$
corrections in this temperature range are probably overestimated. The
3D corrections in $\Te$ appear problematic as well. Indeed, while the
photometric and uncorrected $\logg$ distributions are uniform as a
function of effective temperature, the corrected spectroscopic
distribution now shows an accumulation of objects around 25,000 K, as
well as a depletion of objects near 28,000 K.

However, it is important to stress at this point that these 3D
corrections are available for pure helium atmospheres only, and it is
expected that calculations currently underway, which include the
presence of hydrogen, will improve the results. We defer the rest of
our discussion of these 3D corrections to the end of the next section.

\subsection{Mass Distributions}\label{sect2:mass}

\begin{figure*}[t]
\centering
  \includegraphics[clip=true,angle=270,trim=1cm 2cm 1.5cm 1cm,width=\linewidth]{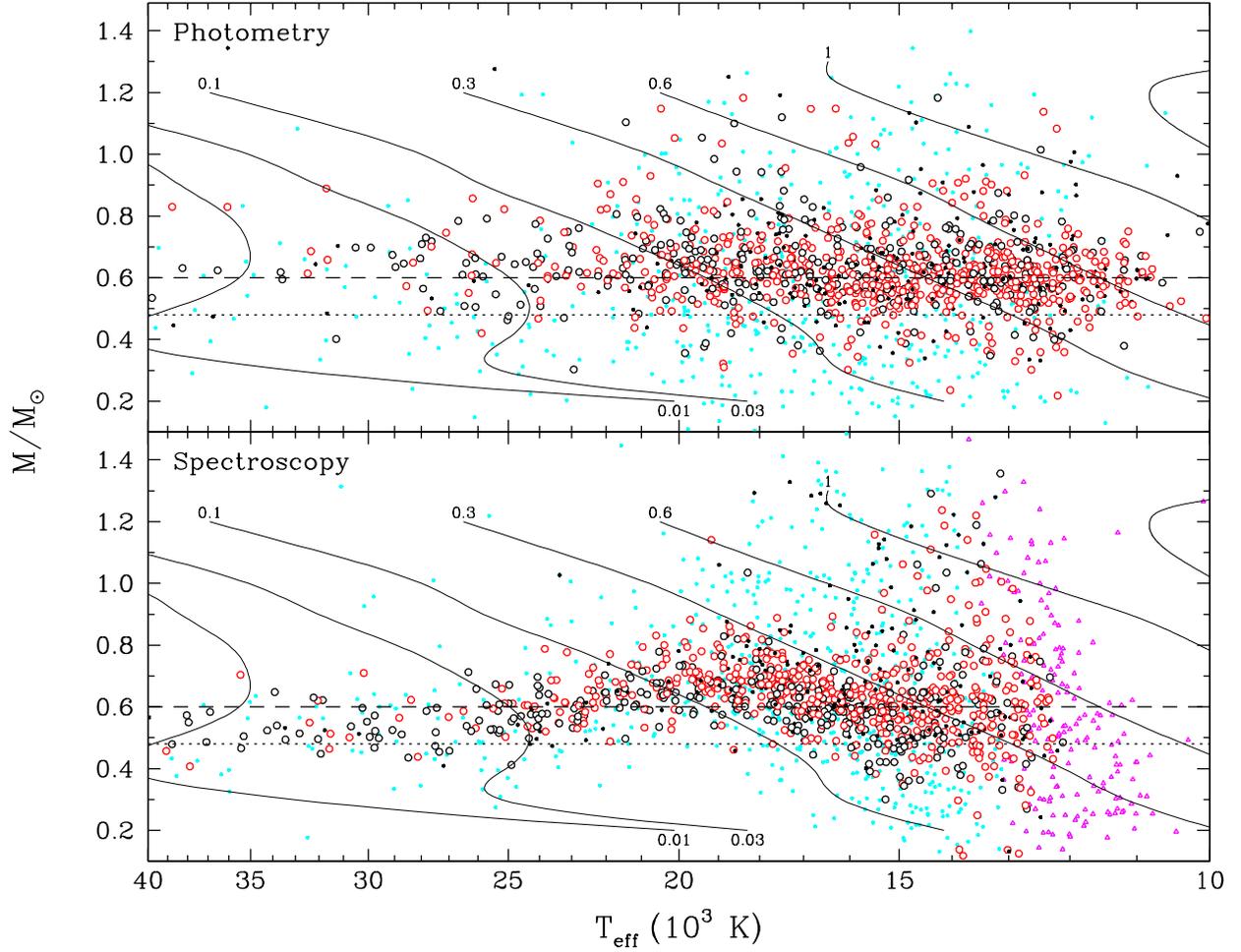}
  \caption{Photometric and spectroscopic masses as a function of
    effective temperature. The description of symbols is identical to
    that of Figure \ref{fig2:logg3D}. Also shown as solid lines are
    theoretical isochrones obtained from models with C/O cores,
    $q(\rm{He})=10^{-2}$, and $q(\rm{H})=10^{-10}$, labeled in units
    of 10$^9$ years. The horizontal dotted and dashed lines are
    located at $M=0.48~\msun$ and $M=0.6~\msun$, respectively (see
    text).}
  \label{fig2:correltm}
\end{figure*}

The stellar radii and $\logg$ values obtained from the photometric and
spectroscopic techniques, respectively, can be converted into stellar
mass using the evolutionary models described in Section
\ref{sect2:theory}. These photometric and spectroscopic masses for all
the DB white dwarfs in our sample are displayed in Figure
\ref{fig2:correltm} as a function of effective temperature. We focus
here on the results from our best data sets, represented by black and
red circles for the DB and DBA white dwarfs, respectively. Not
unexpectedly, we observe here the same behavior as with the $\logg$
distribution. In particular, the photometric mass distribution is well
centered at 0.6 $\msun$, while the spectroscopic distribution exhibits
all the pitfalls previously described. Most noteworthy are the
spectroscopic masses in the $16,000-22,000$ K temperature range, which
are systematically larger than the canonical 0.6 $\msun$ value, while
they appear systematically lower than this value outside this
temperature range. As discussed above, these features can be explained
as a combination of residual flux calibration problems with the SDSS
spectra, 3D hydrodynamical effects, and inadequate van der Waals
broadening, which all affect our spectroscopic solutions.

\begin{figure*}[t]
\centering
  \includegraphics[angle=270,clip=true,trim=3cm 2.5cm 2cm 2.5cm,width=\linewidth]{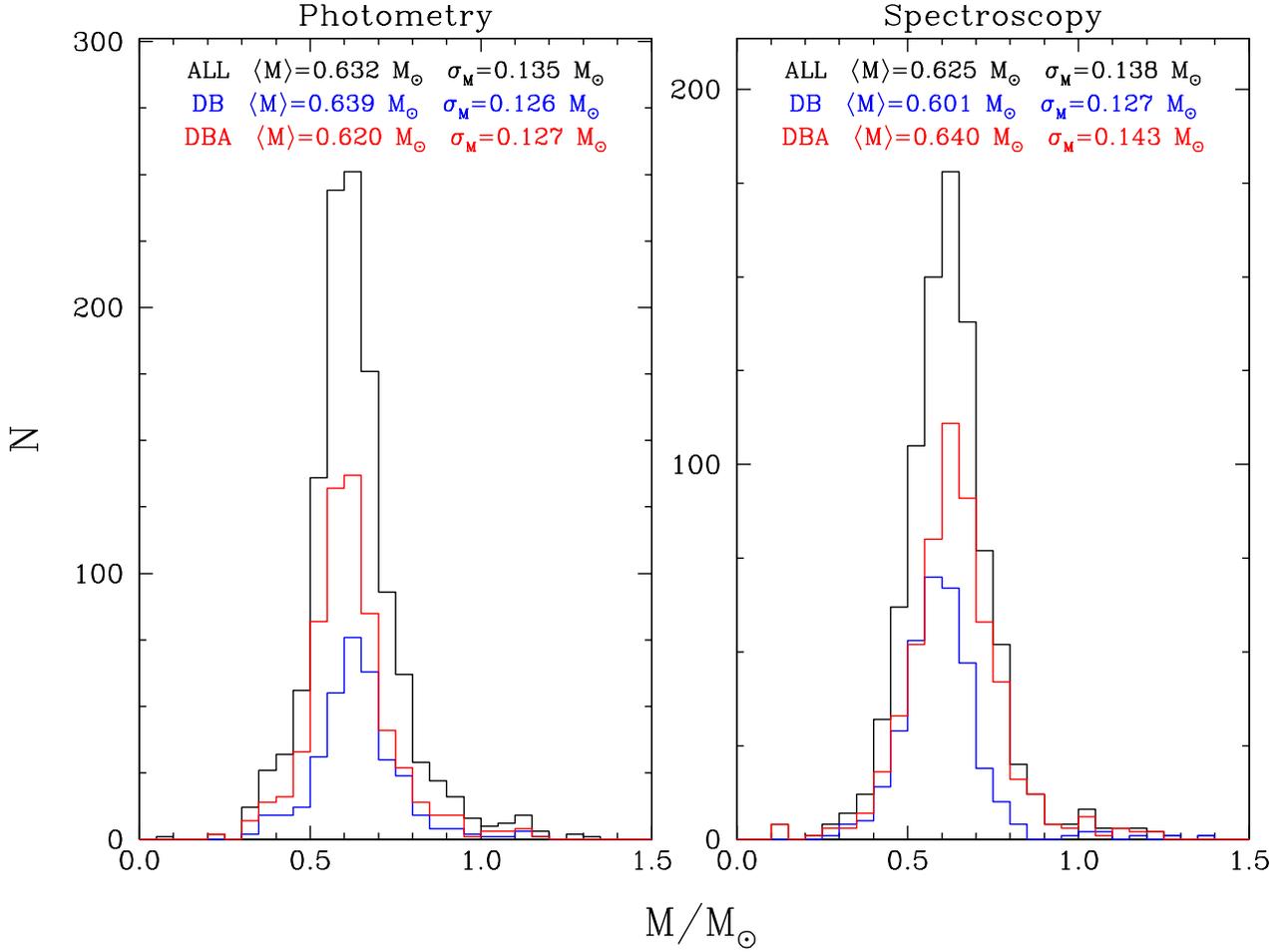}
  \caption{Left: Photometric mass distribution for the objects with
    $\sigma_\pi/\pi<0.25$. Right: Spectroscopic mass distribution for
    the DB spectra with $\sn>10$. The objects with very marginal
    helium lines have been excluded. Also shown are the corresponding
    mass distributions for the DB (blue) and DBA (red) subsamples. The
    mean masses and standard deviations are also given in each panel.}
  \label{fig2:histo_M_DB}
\end{figure*}

Another way to investigate the white dwarf masses is to look at the
relative mass distributions. To ensure the best possible mass
values, we restricted our spectroscopic sample to the objects with
$\sn > 10$, and discarded the spectra showing only marginal helium
lines (see Section \ref{sect2:logg}). Similarly, for the photometric
mass distribution, we restricted our sample to $\sigma_\pi/\pi <
0.25$. The resulting mass distributions, both photometric
and spectroscopic, are displayed in Figure
\ref{fig2:histo_M_DB}. Despite the problems mentioned in the previous
paragraph, the relative mass distributions obtained from photometry
and spectroscopy are remarkably similar. They both have a mean mass of
$\sim$$0.63~\msun$, and very similar dispersions ($\sigma_{M,{\rm
    phot}}=0.135~\msun$ and $\sigma_{M,{\rm
    spec}}=0.138~\msun$). These mean masses are much lower than the
value of $0.706~\msun$ reported by \citet{Koester2015} for their
complete sample, which can probably be attributed to differences in
model atmospheres and fitting techniques (see Section
\ref{sect2:KK2015}). Note, however, that their favored mean mass for
DB white dwarfs is $0.606~\msun$, based exclusively on the objects
between $\Te=16,000$~K and 22,000~K.

One particular feature often reported regarding the mass distribution
of DB white dwarfs is the complete absence of a low-mass tail
(\citealt{Beauchamp1996,Bergeron2011}; GBB19), suggesting that common
envelope evolution scenarios, which are often invoked to explain
low-mass ($\lesssim 0.48~\msun$) DA stars \citep{BSL92}, do not
produce DB white dwarfs. Our spectroscopic mass distribution displayed
in Figure \ref{fig2:histo_M_DB} shows a few DB white dwarfs with such
low masses, but an examination of Figure \ref{fig2:correltm} (where
this low-mass limit is indicated by the dotted line) reveals that
these objects are located either below $\Te=16,000~\K$, where our
solutions are more uncertain due to the improper treatment of van der
Waals broadening, or above $\Te\sim 25,000~\K$, where the calibration
issues with the SDSS spectra affect the spectroscopic solutions. The
photometric mass distributions in both Figures \ref{fig2:correltm} and
\ref{fig2:histo_M_DB}, which are not affected by these problems, also
show several low-mass objects, but these are most likely unresolved
double degenerates, as discussed in Section
\ref{sect2:DB+DX}. Therefore, we find no compelling evidence in our
analysis for the existence of low-mass DB white dwarfs, a conclusion
also reached by \citet{Beauchamp1996}, \citet{Bergeron2011}, and
GBB19.

\begin{figure*}[t]
\centering
  \includegraphics[clip=true, trim=1cm 3cm 1cm 1cm,width=0.8\linewidth]{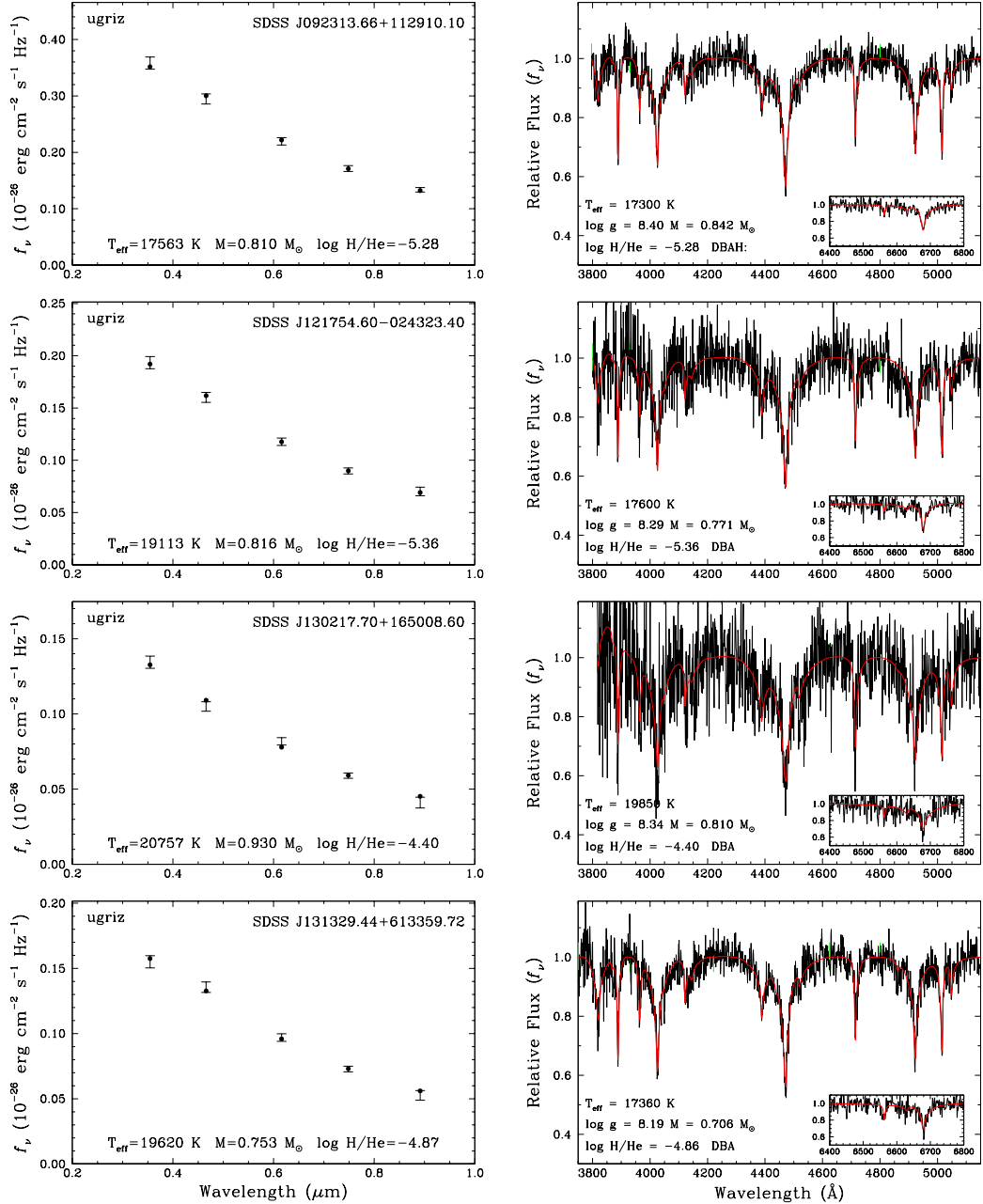}
  \caption{Left: Best photometric fits to four massive DB white
    dwarfs. The error bars represent the observed $ugriz$ magnitudes
    and associated uncertainties, while the filled circles represent
    the best-fit model. Right: Corresponding best spectroscopic
    fits. The best-fit model (red) is plotted over the normalized
    observed spectrum (black). The inset shows the region near
    H$\alpha$ used to determine the hydrogen abundance, or upper
    limits. The derived atmospheric and physical parameters are also
    given.}
  \label{fig2:massive}
\end{figure*}

The high-mass tail of the spectroscopic mass distribution observed in
Figure \ref{fig2:histo_M_DB} is often attributed to the improper
treatment of van der Waals broadening in model atmospheres
(\citealt{Bergeron2011,Koester2015}; GBB19). This is supported by the
fact that most objects with large spectroscopic masses are located
below 16,000 K where this type of line broadening dominates (see
Figure \ref{fig2:correltm}). This might not be the whole story,
however, since the photometric mass distribution also shows a similar
high-mass tail (see also Figure \ref{fig2:correltm}). We present in
Figure \ref{fig2:massive} our best photometric and spectroscopic fits
for four massive DB white dwarfs in our sample. In all cases, the
photometric and spectroscopic solutions are in good agreement, within
the uncertainties. Note also that they are not in the temperature
regime where van der Waals broadening dominates. Consequently, their
large inferred masses appear real. The existence of massive DA white
dwarfs is usually explained by stellar mergers
\citep{Iben1990,Kilic2018}, or as a result of the initial-to-final
mass relation \citep{ElBadry2018}. The same mechanisms can possibly be
invoked as well to explain the presence of such massive DB white
dwarfs in our sample.

Another possible explanation for the origin of massive DB white dwarfs
involves the so-called Hot DQ stars, whose atmospheres are dominated
by carbon \citep{Dufour2007,Dufour2008}. Since Hot DQ white dwarfs are
only found above $\Te \sim 18,000~\K$, \citet{Bergeron2011} proposed
that they somehow transform into DB stars around that temperature,
through a currently unknown physical mechanism. An examination of the
upper panel of Figure \ref{fig2:correltm} actually reveals that
massive DB white dwarfs start to appear below $\Te \sim 22,000~\K$,
coinciding with the coolest Hot DQ stars known today. The hot DQs also
tend to be massive, since they are most likely the end result of white
dwarf mergers \citep{Dunlap}. Therefore, we suggest that some of the
massive DB white dwarfs observed in our sample could be former Hot DQ
stars.

Another particularly important issue is whether the mass distributions
of DB and DBA white dwarfs differ or not. In our analysis, we
considered an object to be a DB star if the spectroscopic technique
could only determine an upper limit on the hydrogen abundance. The
relative photometric mass distributions for DB and DBA white dwarfs
are presented in the left panel of Figure \ref{fig2:histo_M_DB}. The
comparison indicates that the average masses differ by less than 0.02
$\msun$, and that their dispersions are identical, a result that is
readily apparent when looking at the upper panel of Figure
\ref{fig2:correltm}. A similar comparison with the spectroscopic mass
distributions displayed in the right panel of Figure
\ref{fig2:histo_M_DB} suggest that DB white dwarfs are slightly less
massive than DBA stars, by about 0.04 $\msun$. However, it is also
obvious from the results shown in the lower panel of Figure
\ref{fig2:correltm} that these mass differences stem from all the
problems related with the spectroscopic technique across the entire
temperature range, both observational and theoretical, as already
discussed extensively above. We thus conclude that there is no
significant mass difference between the DB and DBA white dwarfs, and
that the explanation for the origin of hydrogen in DBA stars is not
mass related.

\begin{figure}[t]
\centering
  \includegraphics[clip=true,trim=2cm 3cm 2cm 4cm,width=0.8\linewidth]{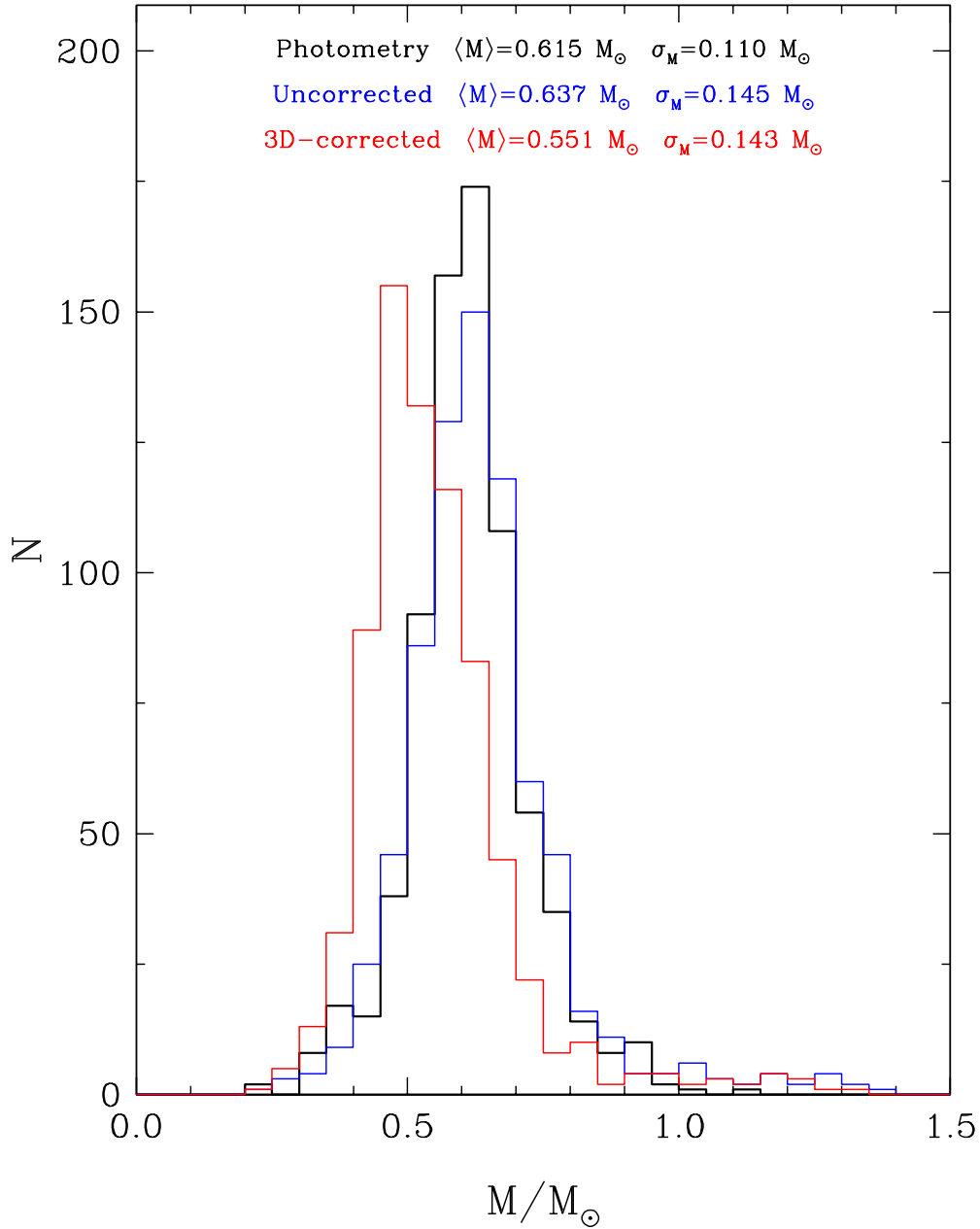}
  \caption{Relative mass distributions obtained from photometry
    (black) and spectroscopy (blue: uncorrected; red: 3D-corrected),
    for the DB white dwarfs in common between the photometric and
    spectroscopic samples. The objects with $\sn<10$,
    $\sigma_\pi/\pi>0.25$, or marginal helium lines have been
    excluded. The mean masses and standard deviations are also given
    in the figure.}
  \label{fig2:histo_corr3D}
\end{figure}

Finally, we go back to our discussion of the 3D hydrodynamical
corrections by looking at the relative mass distributions. In Figure
\ref{fig2:histo_corr3D}, we compare the photometric masses --- which
are unaffected by 3D effects --- with those obtained
spectroscopically, both uncorrected and 3D-corrected. Note that the
mass distributions displayed here include only the DB white dwarfs in
common with both the photometric and spectroscopic samples. Although
the photometric and uncorrected spectroscopic mass distributions
overlap almost perfectly, the 3D-corrected distribution is
significantly shifted towards lower masses. Its mean mass is $\langle
M \rangle = 0.55~\msun$, which is $0.06~\msun$ lower than the value
inferred from photometry, $\langle M \rangle=0.61~\msun$. This again
suggests that the proposed 3D corrections in $\logg$ are too
strong. As previously mentioned, however, the corrections applied here
are appropriate for pure helium models only, and it is possible that
3D hydrodynamical models including traces of hydrogen will yield a
more satisfactory agreement. For the time being, since most objects in
our sample show traces of hydrogen, we will refrain from applying the
3D corrections to our spectroscopic parameters in the remainder of
this study.

\subsection{Hydrogen Abundance Distribution}\label{sect2:Habundance}

\begin{figure*}[t]
\centering
  \includegraphics[angle=270,clip=true,trim=3.5cm 2cm 3.5cm 3cm,width=0.8\linewidth]{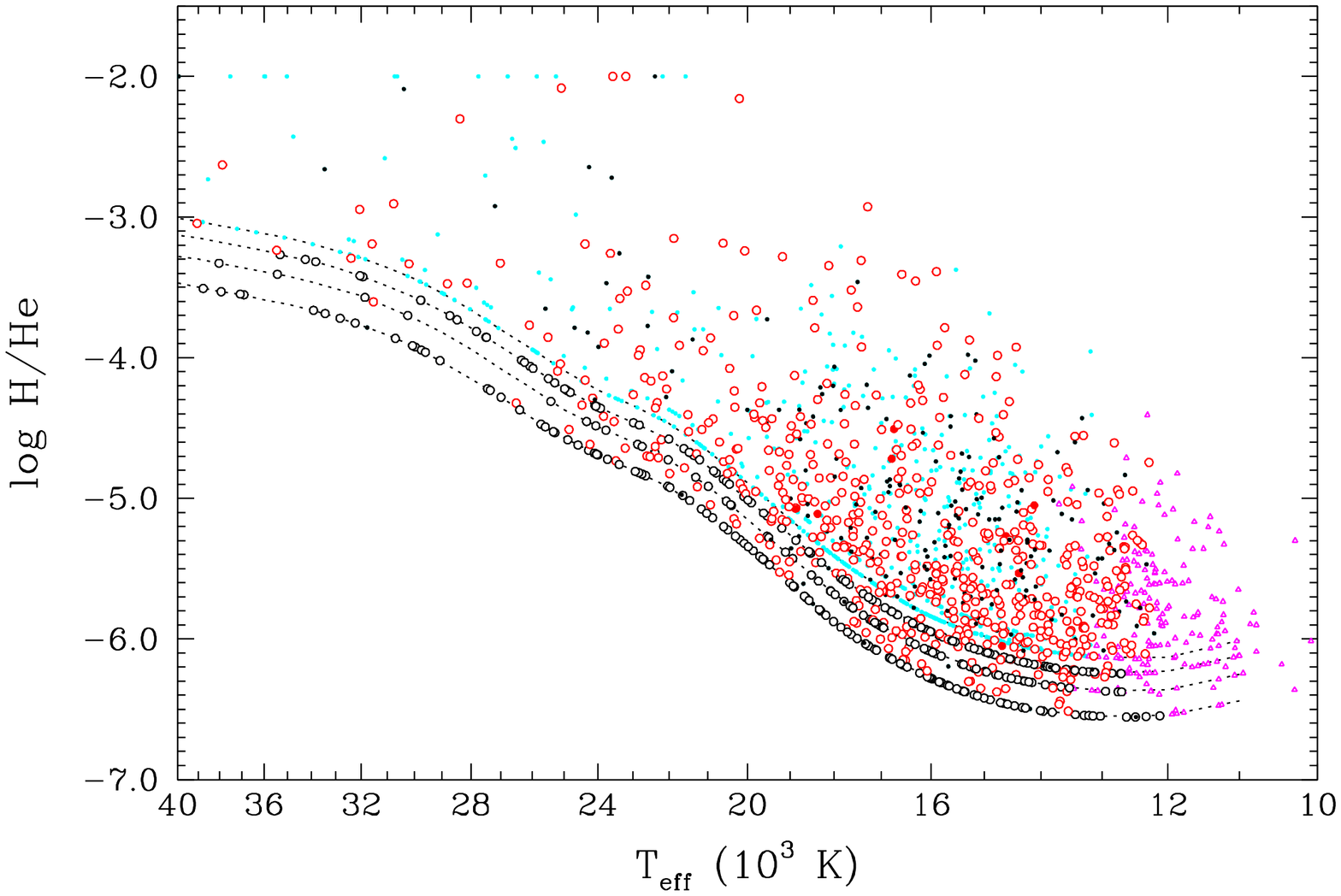}
  \caption{Hydrogen abundances as a function of effective
    temperature. The description of symbols is identical to that of
    Figure \ref{fig2:logg3D}, with the exception that unresolved DB+DA
    candidates (see Section \ref{sect2:DB+DX}) are shown by red filled
    circles. Limits on the hydrogen abundance set by our spectroscopic
    observations at H$\alpha$ are shown as dotted lines for (from
    bottom to top) $\sn > 20$ (200 m\AA\ equivalent width), $15 < \sn
    < 20$ (300 m\AA), $10 < \sn < 15$ (400 m\AA), and $\sn < 10$ (500
    m\AA).  }
  \label{fig2:h_abundance}
\end{figure*}

The final parameter we need to discuss is the hydrogen abundance
ratio, $\nHH$, which is displayed in Figure \ref{fig2:h_abundance} as
a function of effective temperature for all the objects in our
spectroscopic sample. Since the limit of detectability of H$\alpha$
depends on the quality of the spectroscopic observations, we only kept
the spectrum with the highest S/N for the objects with multiple
observations. Also, as mentioned above, we considered an object to be
a DB star if only an upper limit on the hydrogen abundance could be
determined by our fitting technique. With this definition, we find
that 61\% of the objects in our sample are DBA white dwarfs, a ratio
which is similar to the 63\% obtained by \citet{Rolland2018}, but
somewhat lower than the 75\% reported by \citet{Koester2015}, although
their higher value was obtained by restricting their spectroscopic
sample to $\sn > 40$ (their Table 3 actually reveals a much broader
range of values).

The general abundance pattern observed in Figure
\ref{fig2:h_abundance} is consistent with the deepening of the helium
convection zone as the star cools off (see, e.g., Figure 9 of
\citealt{Rolland2018}), in which hydrogen is gradually being diluted
into a larger and more massive convective envelope. It is thus not
surprising to find some of the largest hydrogen abundances at high
temperatures, where the helium convection zone is the shallowest. Note
that in all cases, hydrogen always remains a trace element in the
stellar envelope, and its presence does not affect the structure of
the convection zone in any way \citep{Rolland2018}.

Despite the poor quality of some of the SDSS spectra with S/N < 10
(shown by cyan dots in Figure \ref{fig2:h_abundance}), we can see that
the derived hydrogen abundances overlap perfectly with the bulk of our
other determinations, except at the hot end of the sequence where
H$\alpha$ becomes increasingly more difficult to detect, especially in
low S/N spectra. In such cases our fitting algorithm may yield
unreliable hydrogen abundance measurements. At the cool end of the
sequence, however, we are able to obtain reasonable hydrogen
abundances, even from white dwarf spectra where helium lines are
barely detected.

\begin{figure*}[t]
\centering
  \includegraphics[clip=true, trim=1cm 14cm 1cm 1cm,width=\linewidth]{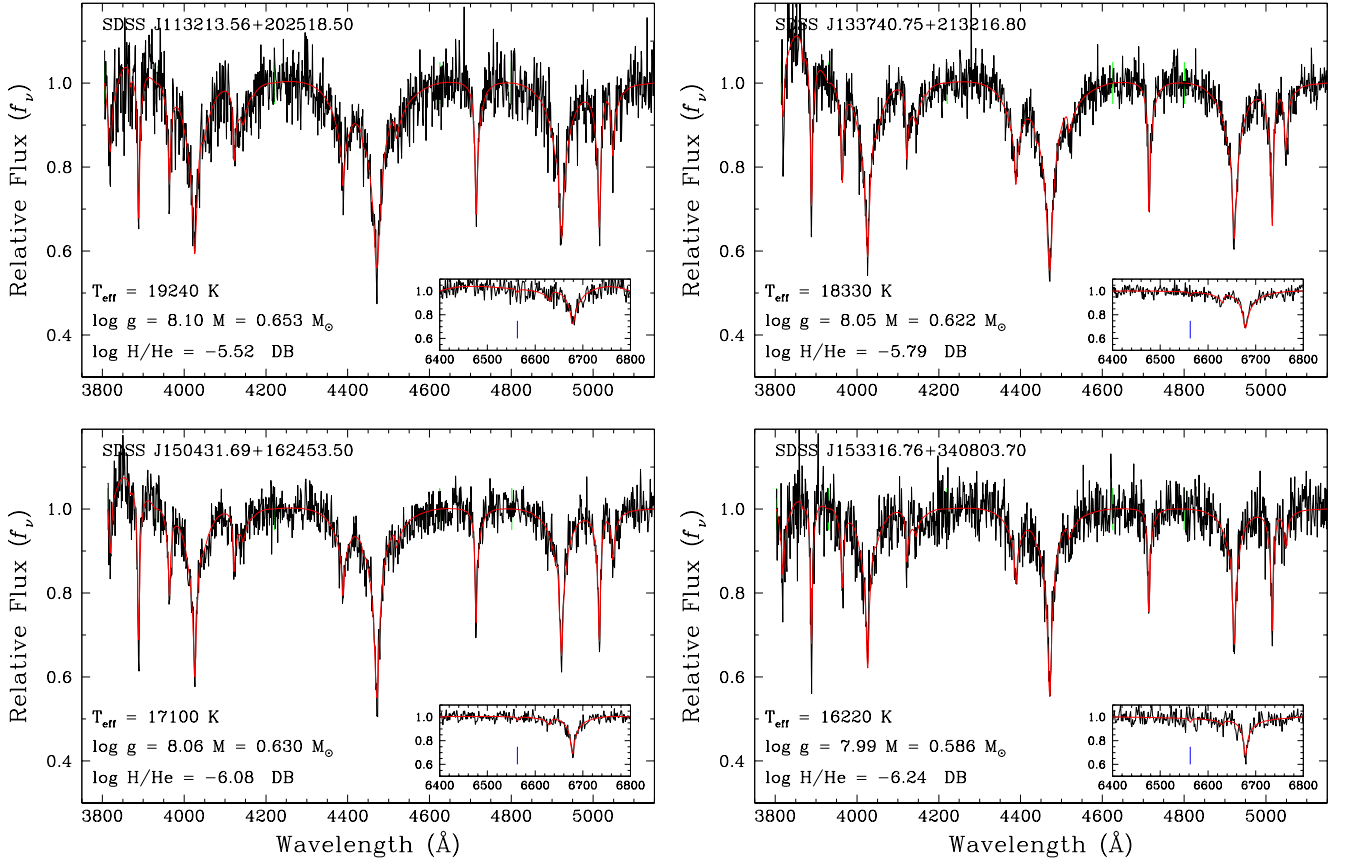}
  \caption{Best spectroscopic fit to four cool DB white dwarfs with no
    detectable H$\alpha$ feature. The best-fit model (red) is plotted
    over the normalized observed spectrum (black). The inset shows the
    region near H$\alpha$ (indicated by the tick mark) used to
    determine the hydrogen abundance, or upper limits. The derived
    atmospheric parameters are also given in each panel.}
  \label{fig2:No_H}
\end{figure*}

Perhaps the most striking feature in Figure \ref{fig2:h_abundance} is
the range of H/He values at a given effective temperature, which can
reach as much as 3 orders of magnitude near $\Te\sim 17,000$~K. Note
that at these temperatures, H$\alpha$ can be easily detected
spectroscopically, and hydrogen abundances as low as $\nHH \sim
10^{-6}$ can be effectively measured, even with a relatively low S/N
spectrum (see the detection limits in Figure
\ref{fig2:h_abundance}). Even so, there is a significant fraction of
DB white dwarfs in our sample, {\it found at all temperatures},
showing no H$\alpha$ absorption feature. Four of these, selected in a
temperature range where H$\alpha$ can be most easily detected, are
displayed in Figure \ref{fig2:No_H}. These cool DB stars have so
little hydrogen that they must have maintained hydrogen-poor envelopes
throughout their evolution. Hence, whatever scenario is invoked to
explain the presence of hydrogen in DBA white dwarfs, it must be able
to account for the large spread in hydrogen abundances observed
here. We discuss this issue at length in Section
\ref{sect2:AtmosphericH}.

 \subsection{Comparison with Koester \& Kepler (2015)}\label{sect2:KK2015}

\begin{figure*}[t]
\centering
  \includegraphics[angle=270,clip=true,trim=2cm 2cm 2.5cm 2cm,width=\linewidth]{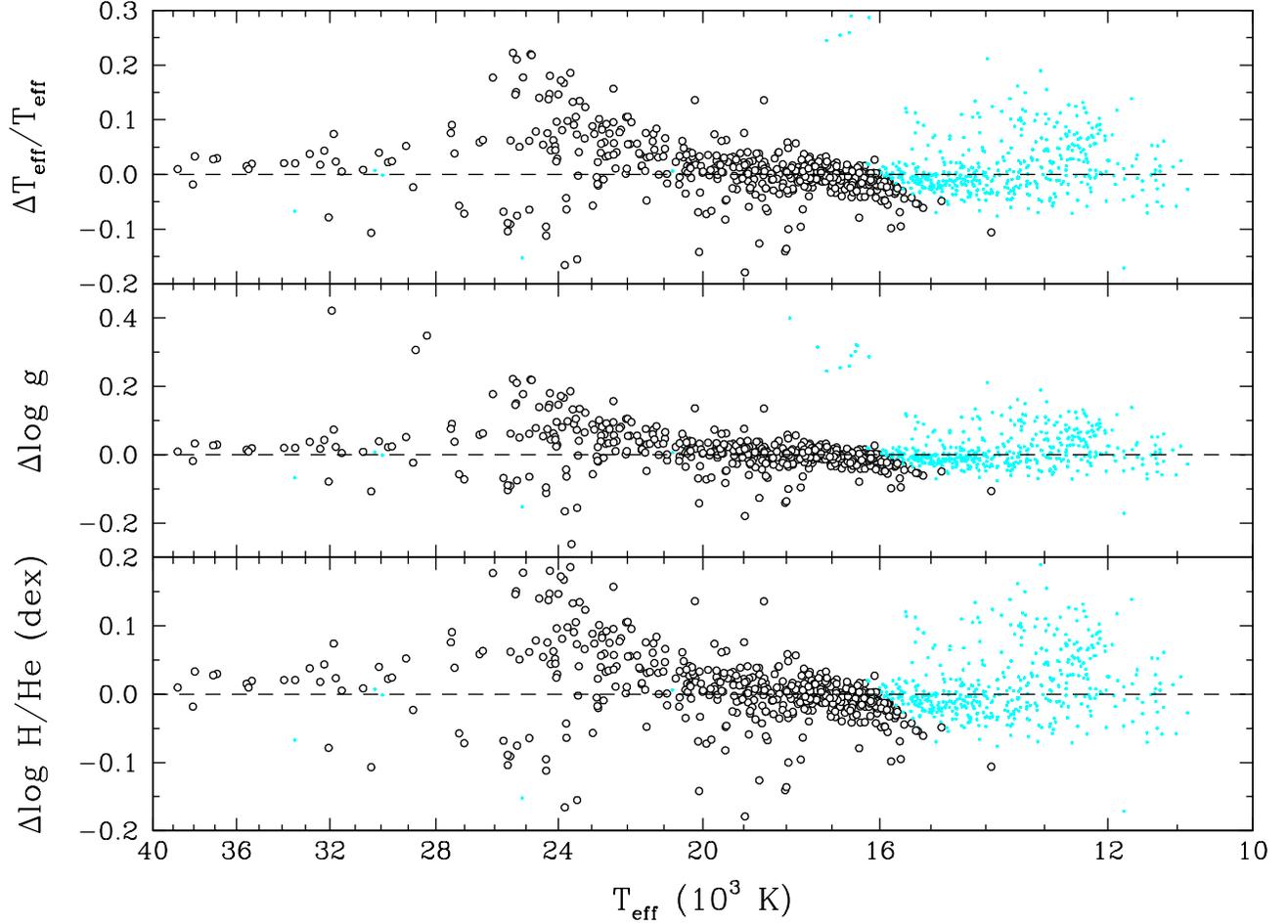}
  \caption{Differences in effective temperatures, surface gravities,
    and hydrogen abundances as a function of effective temperature
    between our analysis and that of \citet{Koester2015}. The cyan
    dots represent the objects for which Koester \& Kepler assumed
    $\logg = 8.0$.}
  \label{fig2:KK15}
\end{figure*}

\citet{Koester2015} performed a similar analysis of the DB white
dwarfs in the SDSS, also drawn from the SDSS DR10 and DR12. The
comparison of their effective temperatures, surface gravities, and
hydrogen abundances for the 996 objects in common with our analysis is
presented in Figure \ref{fig2:KK15}. If we exclude the objects below
$\Te=16,000$~K, for which Koester \& Kepler assumed $\log g = 8.0$, we
find a good overall agreement between our atmospheric parameters and
theirs. There are however some slight discrepancies between the two
sets of values, in particular around $\Te\sim24,000$~K where the
helium lines reach their maximum strength. Around this region, our
effective temperatures, surface gravities, and hydrogen abundances are
in general higher than their values, but there is also a lot more
scatter. Koester \& Kepler also mention a deficiency of objects in the
interval $\Te=24,000 - 26,000$~K (see their Figure 1 and Table 2),
which they attribute to either the assumed convective efficiency, or
to flux calibration issues with the SDSS spectra. While we use the
same convective efficiency (ML2/$\alpha$=1.25) and the same SDSS
spectra, we do not see such a depletion of objects in this temperature
range (see, e.g., Figure \ref{fig2:correltm}). All these small
discrepancies between their analysis and ours can probably be
attributed to differences in model atmospheres and/or fitting
techniques.

\citet{Koester2015} also find that, below $\Te \sim 16,000~\K$, their
spectroscopic $\logg$ values (or masses) increase steadily, forming an
almost continuous distribution (see their Figure 1). In particular,
they find almost no cool DB white dwarfs with $\logg \sim 8.0$ in this
temperature range, in contrast with our surface gravity distribution
displayed in Figure \ref{fig2:logg3D}, which shows a spread rather
than a continuous increase in $\logg$ below $\Te \sim 16,000~\K$ (see
also Figure 6 of \citealt{Rolland2018}). Moreover, we find a
significant number of white dwarfs with normal masses (see Figure
\ref{fig2:difflogg}), or even lower than average. These discrepant
results are most likely due to differences in the treatment of van der
Waals broadening between both sets of model atmospheres. This may also
explain the mean mass of 0.706~$\msun$ obtained by Koester \& Kepler
for their complete sample, while we find a much lower value of
$0.63~\msun$.

\subsection{Accuracy and Precision of the Fitting Techniques}\label{sect2:accuprec}

At this point in our analysis, it is worth reevaluating the accuracy
and the precision of both the photometric and spectroscopic techniques
for determining the physical parameters of DB white dwarfs using the
$ugriz$ photometry and optical spectra from the SDSS. In this context,
the precision refers to the level of agreement of a measurement with
itself when it is repeated several times, while the accuracy refers to
the proximity of the measurement to the true physical value.

On the basis of our best data sets --- i.e.~the spectroscopic sample
with $\sn>10$ and the photometric sample with $\sigma_\pi/\pi<0.25$
--- we conclude that, on average, the spectroscopic technique is more
\textit{precise} than the photometric technique for determining the
effective temperatures of DB white dwarfs ---
$\langle\sigma_{\Te}\rangle = 2.59\%$ from spectroscopy versus 8.72\%
from photometry --- {\it when using the SDSS photometric and
  spectroscopic data}.  At low effective temperatures, however, the
photometric technique becomes as precise as the spectroscopic
technique, if not more ($\langle\sigma_{\Te}\rangle\sim 2\%$ from
photometry at 10,000 K). For the determination of stellar masses, the
errors are also smaller from spectroscopy ($\langle\sigma_{M}\rangle=
0.053~\msun$\footnote{We exclude here the temperature range where van
  der Waals broadening becomes a problem in our model spectra.} from
spectroscopy versus 0.112 $\msun$ from photometry). Of course, the
photometric technique may potentially yield more precise mass
measurements than the spectroscopic technique, regardless of the
temperature range, provided that $\sigma_\pi/\pi$ is small enough (see
section \ref{sect2:error_phot}).

As mentioned in GBB19, the synthetic photometry is less affected by
the input physics of the model atmospheres than the model spectra. For
the DB white dwarfs analyzed here, the photometric mass distribution
was found to be well centered on $M=0.6~\msun$, at all temperatures,
while the spectroscopic distribution deviated from this value between
21,000 K and 17,000 K, as well as below 16,000 K (see Section
\ref{sect2:mass}). However, GBB19 found a good agreement between the
photometric and spectroscopic temperatures, except at the hot end of
the distribution, where the $ugriz$ photometry is in the
Rayleigh-Jeans regime. We thus conclude that, for DB stars, both
fitting techniques have a similar accuracy for the determination of
effective temperatures, but the photometric technique is more accurate
for measuring stellar masses.

Since we will compare in Section \ref{sect2:DBtoDA} the distribution
of DB and DA white dwarfs, we briefly summarize some of the results
from GBB19 regarding the DA stars. GBB19 found that the spectroscopic
temperatures of DA stars were $\sim$10\% higher than the photometric
values for $\Te>14,000$~K, which they attributed to some inaccuracy in
the theory of Stark broadening for hydrogen lines. The spectroscopic
and photometric masses, however, were in very good agreement. We thus
conclude that, for the DA stars, the photometric technique is more
accurate than the spectroscopic technique for the determination of
effective temperatures, but both techniques have a similar accuracy
when it comes to mass determinations. As for the DB white dwarfs,
GBB19 also found that the spectroscopic technique yields more precise
temperature and mass measurements than the photometric technique.

\section{Objects of Particular Astrophysical Interest}\label{sect2:Objects}

Our analysis of the atmospheric and physical parameters of DB white
dwarfs, described in the previous section, has revealed the existence
of several objects of particular astrophysical interest. We discuss
these objects in turn.

\subsection{Double Degenerate Candidates}\label{sect2:DB+DX}

Unresolved double degenerate binaries can be identified by their
extremely low spectroscopic masses (see, e.g., \citealt{BSL92}), or
alternatively, by their overluminosities in Hertzsprung-Russel
diagrams (see Figure 10 of \citealt{BRL97}). In the last case, due to
the presence of two stars in the system, the radius is overestimated,
and thus the photometric mass is underestimated. GBB19 have already
identified several such degenerate binaries in the SDSS data, both
spectroscopically and photometrically.

One type of double degenerate system that can be easily recognized is
those composed of a DA and a DB white dwarf, an excellent example of
which is KUV 02196+2816, analyzed in detail by
\citet{Limoges2009}. The observed spectrum of such systems resembles
that of a DBA white dwarf, but the hydrogen lines are usually
extremely strong and poorly reproduced by single star, homogeneous
models (see Figure 19 of GBB19). We identified a total of 10 DA+DB
unresolved double degenerates in our sample, listed in Table
\ref{tab2:DD_DADB}. It is possible to obtain the effective
temperatures of both components of the system by fitting the spectrum
with a combination of pure hydrogen and helium-rich synthetic
spectra. For simplicity, we assume here $\logg = 8.0$ for both
components of the system, and a pure helium atmosphere for the DB
white dwarf. The photometric\footnote{We simply assume $\logg=8$ if no
  trigonometric parallax measurement is available.} and spectroscopic
solutions obtained for these systems under the assumption of a single
star are reported in Table \ref{tab2:DD_DADB}, together with the
effective temperatures obtained for the individual DA and DB
components; our best fits are also presented in Appendix A. Note that
in all cases where a parallax measurement is available, the
photometric $\logg$ values are significantly lower than those inferred
from spectroscopy.

\begin{figure*}[t]
\centering
  \includegraphics[clip=true, trim=1cm 1.5cm 1cm 1.5cm,width=0.7\linewidth]{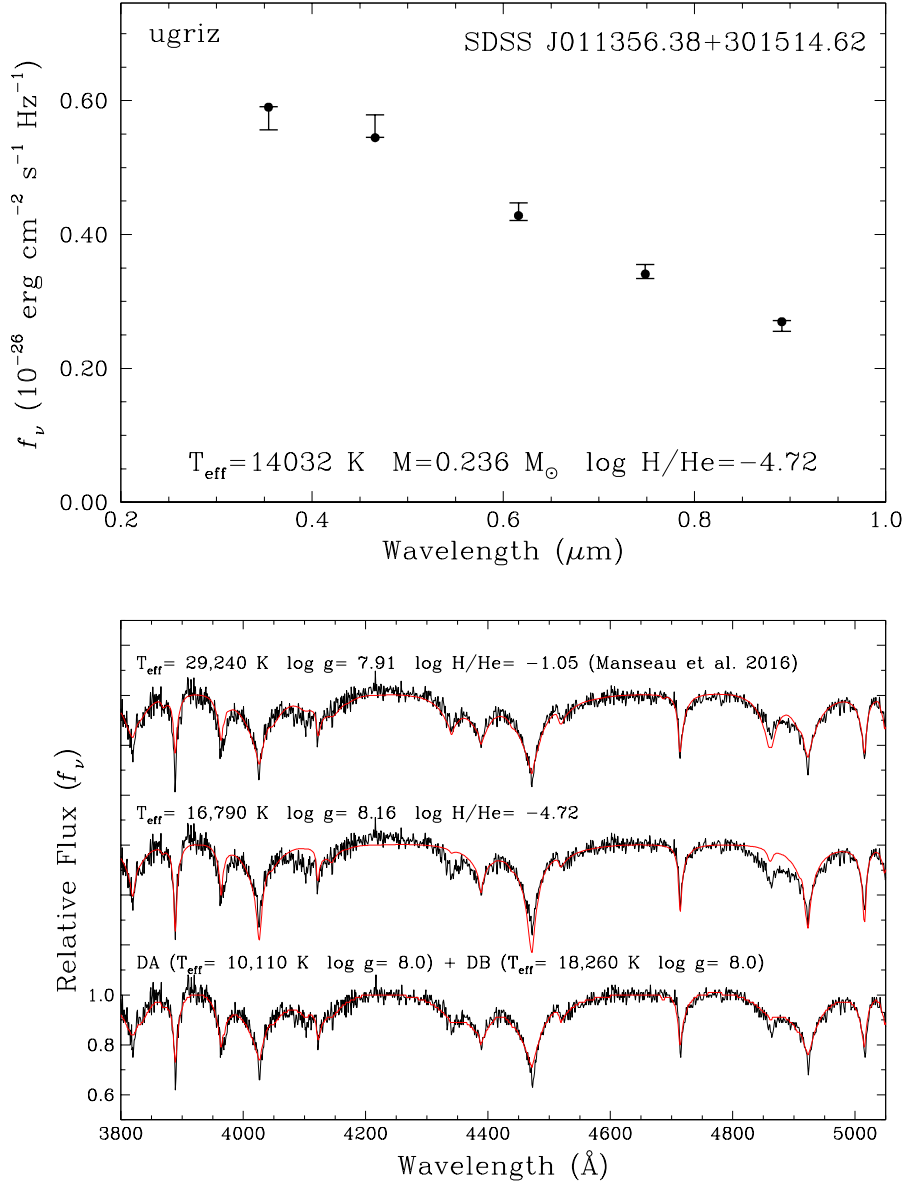}
  \caption{Top panel: Best photometric fit to SDSS
    J011356.38+301514.62 under the assumption of a single DBA white
    dwarf. Bottom panel: (top) best spectroscopic fit obtained by
    \citet{Manseau2016}; (middle) our best spectroscopic fit at lower
    temperature; (bottom) our best fit obtained under the assumption
    of an unresolved DA+DB system. The atmospheric parameters of each
    fit are also given.}
  \label{fig2:J0113+3015}
\end{figure*}

Of these 10 double degenerate systems, SDSS J150506.24+383017.39 has
already been reported by GBB19, while SDSS J011356.38+301514.62 has
been interpreted by \citet{Manseau2016} as a hot, chemically
homogeneous DBA white dwarf with $\Te = 29,200$~K, $\logg=7.91$, and
$\logH=-1.05$. Our photometric solution for this last object,
displayed in the top panel of Figure \ref{fig2:J0113+3015} and
obtained under the assumption of a single star, implies a much lower
temperature. Also, the low inferred photometric mass of only 0.236
$\msun$ clearly indicates the presence of a double degenerate
system. In the lower panel, we compare the spectroscopic fit obtained
by \citet{Manseau2016}, our own spectroscopic solution at lower
effective temperature, and our spectroscopic solution assuming a DA+DB
system. Clearly, this last solution provides not only the best fit to
the observed spectrum, but the average temperature of the system also
agrees perfectly with the photometric temperature. Finally, SDSS
J091016.43$+$210554.20 is classified as magnetic \citep{DR7}, but our
fit displayed in Appendix A indicates that this is
undoubtedly a DA+DB degenerate binary, and interestingly enough, both
components appear to be magnetic! Indeed, Zeeman splitting can be
easily detected in both hydrogen (H$\alpha$ in particular) and helium
lines.

\begin{deluxetable*}{ccccccrc}
\tablecaption{List of DA+DB double degenerate candidates\label{tab2:DD_DADB}}
   \tablehead{ \colhead{} & \multicolumn{2}{c}{Photometry} & \multicolumn{3}{c}{Spectroscopy} & \multicolumn{2}{c}{Deconvolution} \\
	\colhead{SDSS name} & \colhead{$T_{\rm eff}$} & \colhead{$M$} & \colhead{$T_{\rm eff}$} & \colhead{$M$} & \colhead{$\log {\rm H}/{\rm He}$} & \multicolumn{1}{c}{$T_{\rm DA}$} & \colhead{$T_{\rm DB}$} \\
	\colhead{} & \colhead{(K)} & \colhead{($\msun$)} & \colhead{(K)} & \colhead{($\msun$)} & \colhead{} & \multicolumn{1}{c}{(K)} & \colhead{(K)}}  
   \startdata
	011356.38+301514.62 & 14,030 & 0.24 & 16,780 & 0.69 & $-$4.719 & 10,110 & 18,260  \\
	074419.82+302203.40\tablenotemark{a} & 12,670  & 0.59 & 14,610 & 0.62 & $-$5.270 & 11,330 & 15,770 \\
	084716.21+484220.40 & 15,280 & 0.62 & 14,680 & 0.88 & $-$6.050 & 10,080 & 15,470  \\
	091016.43+210554.20\tablenotemark{b} & 14,900 & 0.67 & 15,390 & 0.72 & $-$5.745 & 9,580 & 15,610 \\
	101316.02+075915.20 & 18,880 & 0.32 & 18,870 & 0.73 & $-$5.075 & 12,210 & 30,360 \\
	103609.48+193841.14\tablenotemark{a} & 15,910 & 0.59 & 16,740 & 0.73 & $-$4.509 & 9980 & 16,880 \\
	112711.72+325229.70 & 13,070 & 0.21 & 14,380 & 0.57 & $-$5.532 & 10,410 & 15,260 \\
	113623.54+320403.80\tablenotemark{a} &  15,590 & 0.59 & 18,370 & 0.82 & $-$5.111 & 14,120 & 25,800 \\
	140615.80+562725.90 & 38,700 & 0.45 & 14,340 & 0.91 & $-$5.512 & 13,360 & 48,810 \\
	150506.24+383017.39 & 12,620 & 0.30 & 14,110 & 0.39 & $-$5.049 & 10,180 & 15,490 \\
        \enddata
        \tablenotetext{a}{Photometric solution obtained by assuming $\logg=8.0$.}
        \tablenotetext{b}{The DB and the DA components are both magnetic.}
\end{deluxetable*}

Another type of double degenerate system we found in our sample, also
discussed in GBB19, is composed of two DB white dwarfs. Unlike the
DA+DB systems, these DB+DB binaries cannot be easily recognized from
spectroscopy alone, since the combination of two DB spectra resembles
that of a single DB white dwarf with intermediate atmospheric
parameters (see Figure 20 of GBB19). However, they can still be
identified by comparing the spectroscopic and photometric masses
since, as discussed above, unresolved double degenerate binaries will
have low inferred photometric masses, but more normal spectroscopic
masses. Because our mass estimates depend on the quality of the data,
we restricted our photometric sample to objects with $\sigma_\pi/\pi <
25\%$, and our spectroscopic sample to objects with $\sn>10$; we also
excluded spectra showing only marginal helium lines. We then flagged
the objects for which $M_{\rm spec}-M_{\rm phot} \geq
0.2~\msun$. Similarly, we also flagged all objects with $M_{\rm phot}
\leq 0.45~\msun$, because single star evolution predicts that such
low-mass white dwarfs could not have formed within the lifetime of the
Galaxy. After removing the previously identified DA+DB systems, we
were left with 55 DB+DB unresolved double degenerate candidates. These
systems, as well as the best photometric and spectroscopic solutions,
are listed in Table \ref{tab2:DD_DB}. For the DA+DB systems, we were
able to separate the contributions of each white dwarf. In the case of
DB+DB binaries, it is possible, in principle, to deconvolve the
parameters of both systems using the photometry and parallax
information \citep{Bedard2017}, but this is clearly outside the scope
of this paper. We would like to note that, since we applied a
restriction on the quality of the trigonometric parallax as well as on
$\sn$, the list of DB+DB double degenerate systems given in Table
\ref{tab2:DD_DB} is by no means complete.

\startlongtable
\begin{deluxetable*}{cccccc||cccccc}

\tabletypesize{\scriptsize}

\tablecaption{List of DB+DB double degenerate candidates\label{tab2:DD_DB}}
      \tablehead{ \colhead{}  & \twocolhead{Photometry} & \multicolumn{3}{c}{Spectroscopy} & \colhead{} & \twocolhead{Photometry} & \multicolumn{3}{c}{Spectroscopy} \\
	\colhead{SDSS name} & \colhead{$T_{\rm eff}$} & \colhead{$M$} & \colhead{$T_{\rm eff}$} & \colhead{$M$} & \colhead{$\log {\rm H}/{\rm He}$} & \colhead{SDSS name} & \colhead{$T_{\rm eff}$} & \colhead{$M$} & \colhead{$T_{\rm eff}$} & \colhead{$M$} & \colhead{$\log {\rm H}/{\rm He}$} \\
	\colhead{} & \colhead{(K)} & \colhead{($\msun$)} & \colhead{(K)} & \colhead{($\msun$)} & \colhead{} & \colhead{} & \colhead{(K)} & \colhead{($\msun$)}& \colhead{(K)} & \colhead{($\msun$)} & \colhead{}} 

\startdata
000730.75+275111.90 &  13,932 &    0.41 &  16,305 &    0.63 & $ -5.880$ & 120203.13+285647.07 &  16,043 &    0.38 &  17,179 &    0.69 & $ -5.231$ \\
002153.33+083141.82 &  27,313 &    0.49 &  17,789 &    0.81 & $ -5.733$ & 120735.19+225905.70 &  25,020 &    0.46 &  20,592 &    0.68 & $ -5.201$\\
004900.48-094203.00 &  18,466 &    0.46 &  19,419 &    0.78 & $ -4.434$ & 122444.73+174145.85\tablenotemark{a} &  15,967 &    0.40 &  14,527 &    0.61 & $ -5.613$\\
010532.40+064234.18\tablenotemark{a} &  13,368 &    0.36 &  13,404 &    0.41 & $ -5.694$ & 123230.41+035036.70\tablenotemark{a} &  19,349 &    0.38 &  18,295 &    0.57 & $ -5.801$ \\
011023.82+223716.25\tablenotemark{a} &  12,275 &    0.42 &  12,771 &    0.59 & $ -5.714$ & 123735.52+602833.00\tablenotemark{a} &  12,195 &    0.22 &  15,863 &    0.48 & $ -5.649$\\
011409.86+272739.42 &  15,833 &    0.34 &  17,540 &    0.68 & $ -4.887$ & 124058.65+532623.60 &  16,327 &    0.57 &  17,591 &    0.89 & $ -5.271$ \\
020409.84+212948.58 &  15,071 &    0.59 &  20,605 &    0.83 & $ -3.186$ & 125030.21+594932.90\tablenotemark{a} &  14,844 &    0.42 &  15,831 &    0.59 & $ -5.963$\\
024232.63-050954.75\tablenotemark{a} &  12,181 &    0.40 &  14,438 &    0.56 & $ -5.871$ & 130106.26+023455.30 &  17,660 &    0.45 &  17,706 &    0.77 & $ -5.570$\\
034741.96+010823.80 &  26,320 &    0.55 &  17,809 &    0.78 & $ -4.844$ & 130830.53+470017.90\tablenotemark{a} &  14,705 &    0.42 &  17,109 &    0.54 & $ -5.737$\\
052941.58+603806.80\tablenotemark{a} &  13,612 &    0.39 &  16,823 &    0.50 & $ -5.665$ & 131658.16+305148.00 &  18,987 &    0.52 &  19,635 &    0.76 & $ -4.448$ \\
064452.30+371144.30\tablenotemark{a} &  14,077 &    0.41 &  15,241 &    0.57 & $ -6.030$ & 141337.74+450431.60 &  19,841 &    0.36 &  24,983 &    0.61 & $ -4.222$\\
075224.32+150352.34\tablenotemark{a} &  12,219 &    0.38 &  13,562 &    0.54 & $ -6.373$ & 141621.79+322638.60\tablenotemark{a} &  31,289 &    0.40 &  35,428 &    0.47 & $ -3.407$ \\
082323.20+360834.79\tablenotemark{a} &  14,876 &    0.39 &  15,180 &    0.65 & $ -6.208$ & 144650.87+285142.30 &  22,944 &    0.30 &  22,817 &    0.57 & $ -4.479$\\
083024.17+455206.02\tablenotemark{a} &  15,272 &    0.45 &  14,546 &    0.57 & $ -5.296$ & 150301.95+053414.05\tablenotemark{a} &  13,681 &    0.45 &  13,120 &    0.75 & $ -5.705$ \\
093512.70+003857.12\tablenotemark{a} &  12,472 &    0.36 &  12,853 &    0.60 & $ -6.053$ & 150647.60+310313.30\tablenotemark{a} &  18,682 &    0.42 &  18,530 &    0.59 & $ -5.388$ \\
093806.30+032242.53 &  20,190 &    0.49 &  18,864 &    0.74 & $ -5.068$ & 152320.96+005525.10\tablenotemark{a} &  13,317 &    0.35 &  13,758 &    0.78 & $ -6.207$\\
094023.58+185837.24\tablenotemark{a} &  12,891 &    0.33 &  13,449 &    0.53 & $ -6.219$ & 153024.23+331549.72 &  18,154 &    0.47 &  17,544 &    0.70 & $ -4.340$ \\
094638.77+621759.50\tablenotemark{a} &  12,248 &    0.40 &  12,840 &    0.34 & $ -6.244$ & 153316.76+340803.70 &  16,636 &    0.36 &  16,223 &    0.58 & $ -6.243$\\
095455.12+440330.30 &  18,398 &    0.59 &  23,829 &    0.79 & $ -4.416$ & 153735.17+063848.07 &  20,498 &    0.49 &  19,785 &    0.79 & $ -3.663$ \\
100140.17+025853.19 &  16,397 &    0.52 &  18,526 &    0.73 & $ -4.689$ & 154811.34+083613.21 &  21,175 &    0.47 &  19,480 &    0.70 & $ -4.802$ \\
100904.42+060817.50 &  18,565 &    0.45 &  18,325 &    0.83 & $ -5.384$ & 161735.37+311645.41 &  18,860 &    0.31 &  18,470 &    0.60 & $ -3.593$ \\
101022.37+272239.30 &  17,118 &    0.38 &  16,028 &    0.59 & $ -6.085$ & 165339.17+174838.84 &  16,633 &    0.41 &  18,714 &    0.70 & $ -4.916$\\
101249.63+412311.04 &  15,888 &    0.52 &  16,363 &    0.89 & $ -5.260$ & 165946.51+393418.30\tablenotemark{a} &  19,311 &    0.39 &  24,325 &    0.50 & $ -4.454$ \\
102953.32+020812.45\tablenotemark{a} &  14,963 &    0.39 &  14,201 &    0.70 & $ -4.853$ & 172243.19+603059.70 &  16,614 &    0.49 &  18,921 &    0.77 & $ -5.622$ \\
103033.20+385447.59 &  21,194 &    0.44 &  20,387 &    0.80 & $ -4.810$ & 231041.15+141600.80\tablenotemark{a} &  15,781 &    0.42 &  14,938 &    0.49 & $ -6.097$ \\
104117.42+231036.40 &  25,880 &    0.42 &  21,596 &    0.63 & $ -4.784$ & 232344.88+150858.80 &  14,800 &    0.33 &  19,707 &    0.88 & $ -4.321$\\
105829.24+655227.20\tablenotemark{a} &  13,665 &    0.43 &  15,381 &    0.42 & $ -6.182$ & 233305.10+005155.90 &  17,239 &    0.35 &  21,506 &    0.68 & $ -4.405$\\
111946.75+673631.10\tablenotemark{a} &  13,284 &    0.39 &  15,207 &    0.60 & $ -5.625$ & &  & &  &  & \\
\enddata
\tablenotetext{a}{Based on photometry only}
\end{deluxetable*}

\begin{figure*}[t]
\centering
  \includegraphics[clip=true,angle=270,trim=1cm 2cm 1.5cm 1cm,width=\linewidth]{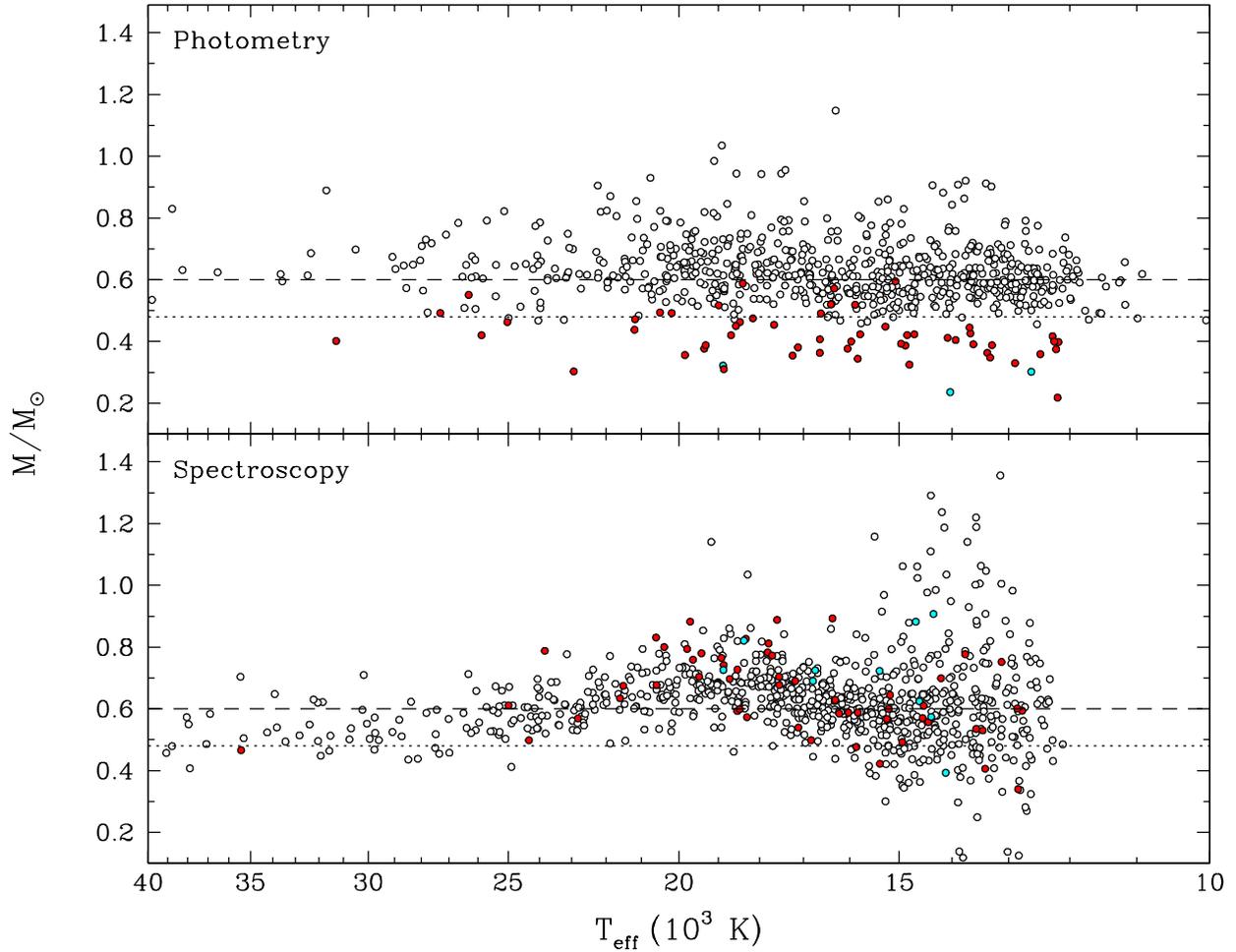}
  \caption{Photometric (top) and spectroscopic (bottom) masses as a
    function of effective temperature. DB+DB and DA+DB double
    degenerate candidates are shown as red and cyan circles,
    respectively. The horizontal dotted and dashed lines are located
    at $M=0.48~\msun$ and $M=0.6~\msun$, respectively.}
  \label{fig2:correltm_DD}
\end{figure*}

We summarize the results of this section by showing in Figure
\ref{fig2:correltm_DD} the location of both DA+DB and DB+DB double
degenerate candidates in a mass versus $\Te$ diagram. Remember that
the photometric and spectroscopic temperatures generally differ for
these systems. One can see that both types of binary systems are
impossible to detect in spectroscopy alone, in contrast with the case
of DA stars \citep[see, e.g.,][]{BSL92}. However, most binary systems
--- but not all of them --- appear as low-mass white dwarfs in the
photometric mass distribution, even the DA+DB binaries.

\subsection{DBA White Dwarfs with Large Hydrogen Abundances}\label{sect2:DBA}

\begin{figure*}[t]
\centering
  \includegraphics[clip=true, trim=1cm 3cm 1cm 1cm,width=0.83\linewidth]{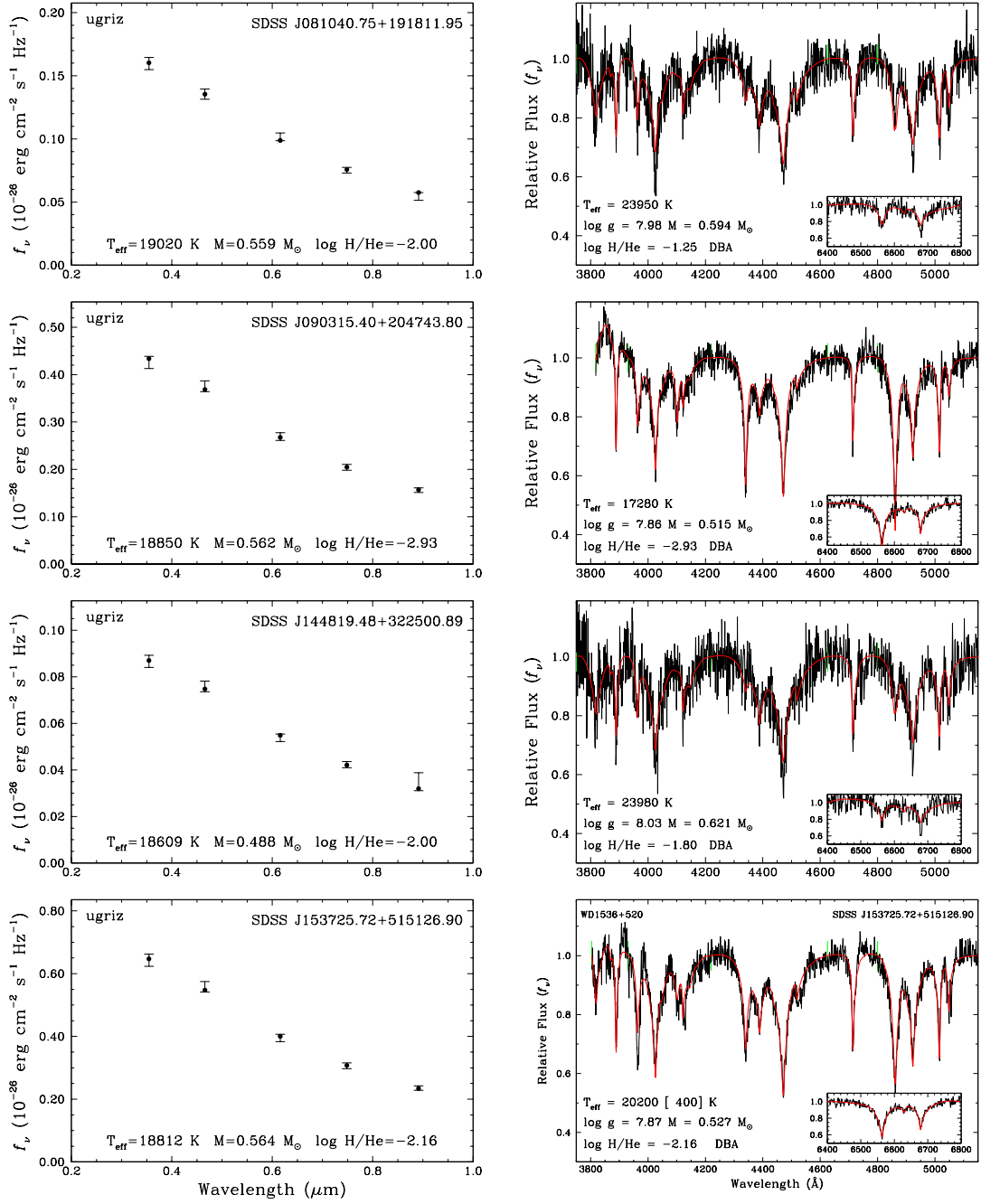}
  \caption{Best photometric (left) and spectroscopic (right) fits to
    four DBA white dwarfs in our sample with extremely large hydrogen
    abundances. The display is similar to that described in Figure
    \ref{fig2:massive}.}
  \label{fig2:highH}
\end{figure*}

There are several DBA white dwarfs in our sample with extremely large
hydrogen abundances, defined arbitrarily here as $\log \nHH>-3$. While
we can find 9 such objects in Figure \ref{fig2:h_abundance}, only a
single white dwarf with a very large hydrogen abundance ($\log
\nHH\sim-2$) has been identified by \citet[][see their Figure
  5]{Rolland2018}, namely PG 1311+129, also discussed at length by
\citet{Bergeron2011}. Among these 9 DBA stars in our sample, the five
hottest objects above $\Te=24,000$~K have spectra with $\sn\sim10$ and
the presence of H$\alpha$ cannot be confirmed with certainty. However,
the four cooler DBA white dwarfs, displayed in Figure
\ref{fig2:highH}, have strong, and well-defined H$\alpha$
features. One of these objects is SDSS J153725.72+51526.90 (WD
1536+520), also analyzed in detail by \citet{Farihi2016}. Our best
spectroscopic fit yields a hydrogen abundance of $\logH=-2.16$, while
Farihi et al.~reported an even larger hydrogen abundance of
$\logH=-1.70$, as well as large abundances of various heavy elements
(O, Mg, Al, Si, Ca, Ti, Cr, Fe). Farihi et al.~concluded that WD
1536+520 was currently accreting debris from a rocky and H$_2$O-rich
parent body. The accretion of hydrogen in this process would also be
responsible for the abnormally large hydrogen abundances observed in
this DBA(Z) white dwarf. We discuss these objects further in Section
\ref{sect2:AtmosphericH}.

\subsection{DBZ White Dwarfs}\label{sect2:DBZ}

We found a total of 118 white dwarfs in our sample with a published
spectral type indicating the presence of metals (DBZ/DBAZ) in their
spectrum. This is obviously a lower limit to the true number of DBZ
stars, because the detection of metals --- mostly the Ca \textsc{ii} H
and K doublet --- depends strongly on the spectral resolution and
S/N. For instance, SDSS J153725.72+51526.90 (WD 1536+520), already
displayed in Figure \ref{fig2:highH}, shows large abundances of
various heavy elements (O, Mg, Al, Si, Ca, Ti, Cr, Fe;
\citealt{Farihi2016}), but it is classified as a DBA star in the
SDSS. As discussed in Section \ref{sect2:theory}, the DBZ white dwarfs
in our sample have been fitted with a specific grid of model
atmospheres that includes only calcium in the equation of state and
opacity calculations. This is a simplistic approach, but at least it
has the benefits of including the most important (probably only)
metallic features detected in the SDSS spectra of DBZ white dwarfs.

\begin{figure*}[!t]
    \centering
    \includegraphics[clip=true, trim=1cm 14cm 1cm 1cm,width=\linewidth]{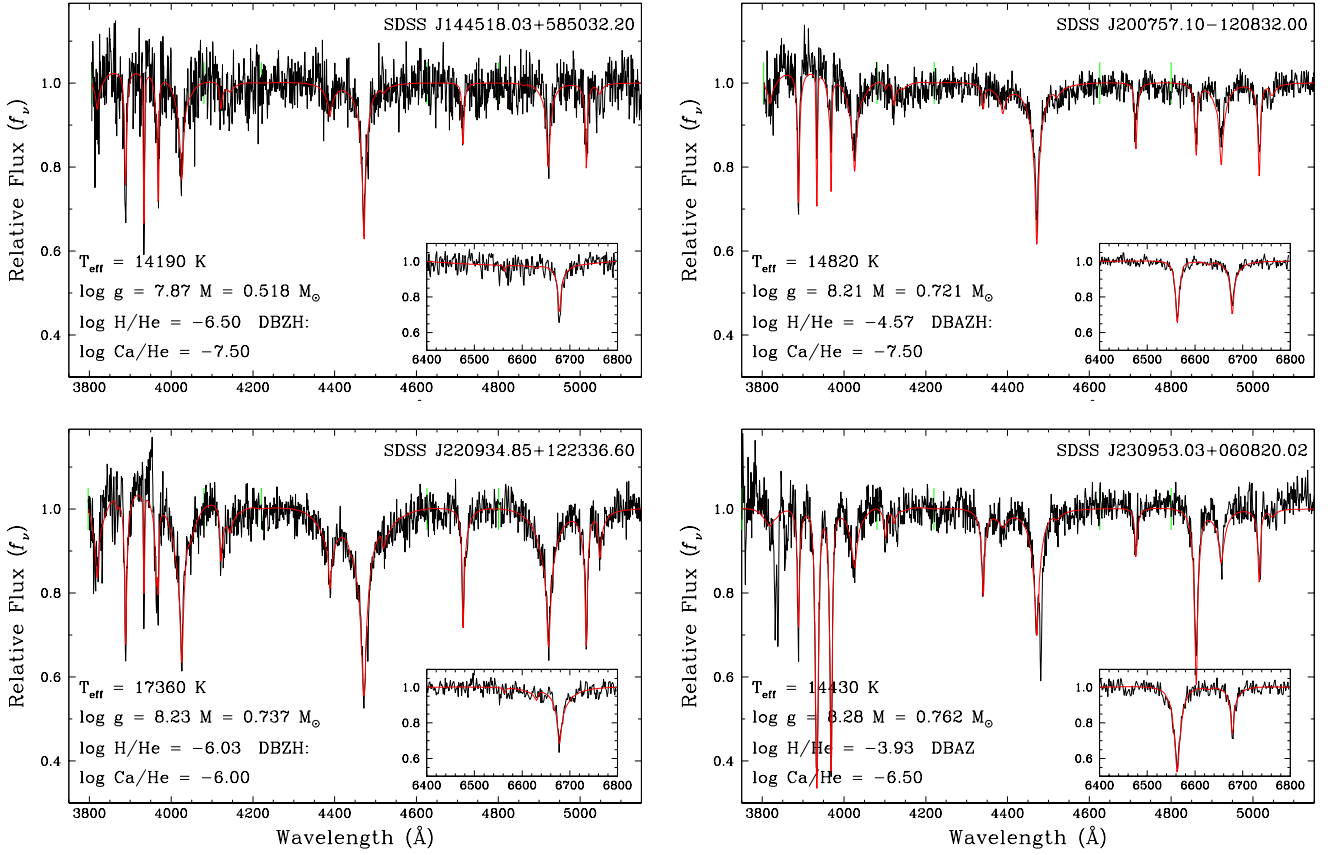}
    \caption{Best spectroscopic fits to four DBZ white dwarfs in our
      sample with strong Ca \textsc{ii} H \& K lines, without (left)
      and with (right) detectable hydrogen features (H$\alpha$ region
      shown in the inset).}
    \label{fig2:Ca650}
\end{figure*}

We present in Figure \ref{fig2:Ca650} our best spectroscopic fits to
four DBZ white dwarfs in our sample with strong Ca \textsc{ii} H and K
lines, two of which also show a detectable H$\alpha$ absorption
feature. In the other two cases, we detect no hydrogen, and its
abundance is thus set to our limit of detectability ($\log
\nHH\ \lesssim -6$). Hence, even though the accretion of metals is
often associated with the probable accretion of water-rich material
(see \citealt{Farihi2016}, \citealt{Gentile2017}, and references
therein), giving rise to large photospheric hydrogen abundances such
as in WD 1536+520, we do find in our sample some objects with large
metal abundances, {\it but with no detectable hydrogen in their
  atmospheres}. This conclusion is based on DBZ white dwarfs found in
a temperature range ($\Te\sim14,000-17,000$~K) where our limit of
detectability is extremely low (see Figure \ref{fig2:h_abundance}). We
come back to this point below.

\subsection{Magnetic White Dwarfs}\label{sect2:magnetic}

In section \ref{sect2:mass}, we discussed the presence of massive DB
white dwarfs in our photometric sample. While most of these were found
below $\Te\sim 22,000$~K, there are two very massive objects located
around $\Te\sim25,500$~K and 36,000 K (small black dots in the upper
panel of Figure \ref{fig2:correltm}). These correspond to SDSS
J094209.49+540157.50 and SDSS J143739.13+315248.80, which are
classified as magnetic DB white dwarfs (DBH), with no clear absorption
lines in their spectra, most likely due to the presence of a strong
magnetic field. For both of these objects, the photometric fit yields
a very large mass of $M\sim 1.3~\msun$. Even though our solutions for
these objects are uncertain at best, magnetic white dwarfs do tend to
be more massive than the non-magnetic population
\citep{Liebert1988,Ferrario2015}, in agreement with our results.

\begin{figure*}[t]
\centering
  \includegraphics[clip=true, trim=1cm 3cm 1cm 1cm,width=0.82\linewidth]{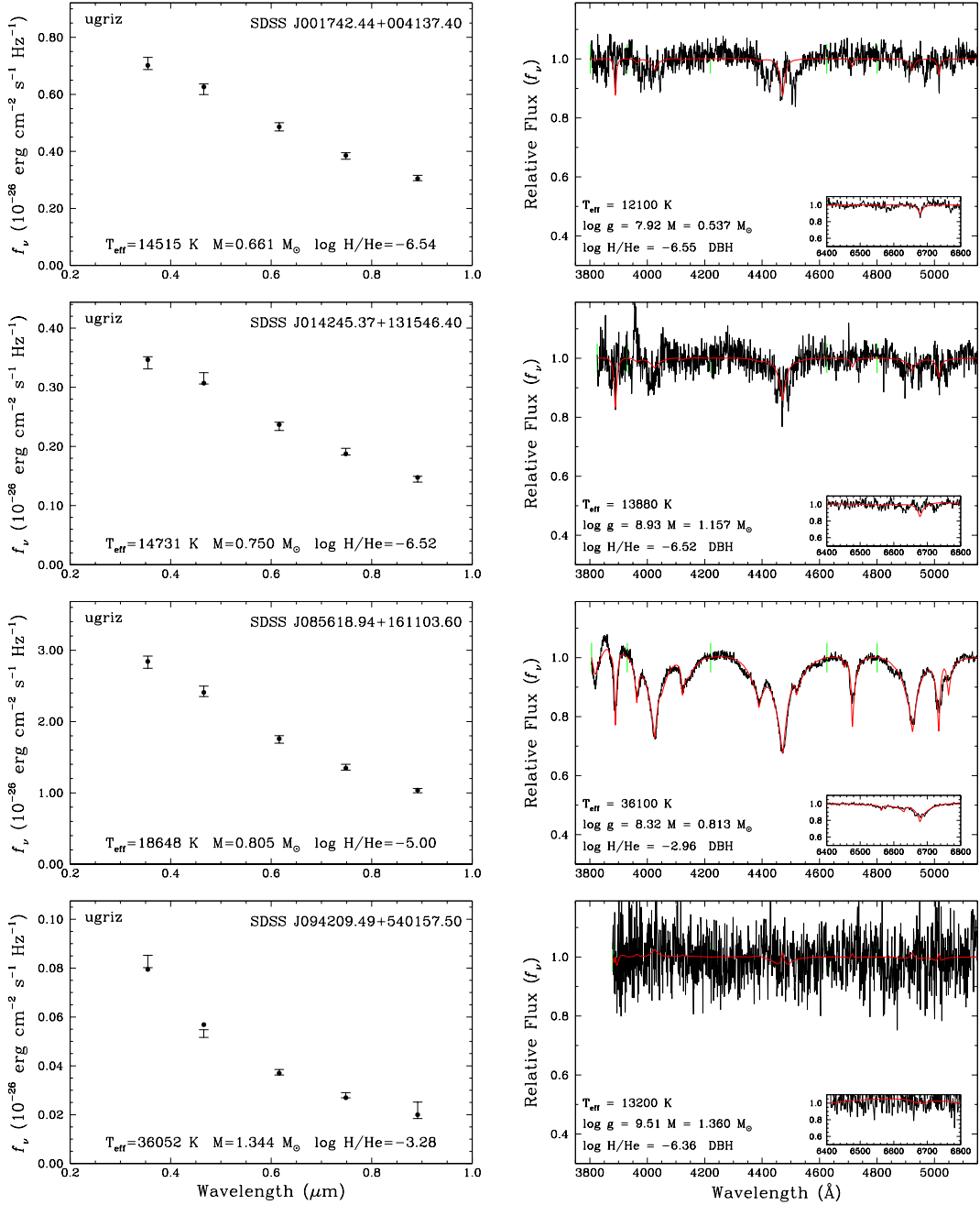}
  \caption{Best photometric (left) and spectroscopic (right) fits to
    four magnetic DB white dwarfs in our sample. The display is
    similar to that described in Figure \ref{fig2:massive}.}
  \label{fig2:magnetic}
\end{figure*}

In some other magnetic DB stars, the magnetic field is strong enough
that Zeeman splitting of the helium lines becomes clearly
visible. Three examples are presented in Figure \ref{fig2:magnetic},
together with our fits to SDSS J094209.49+540157.50, mentioned in the
previous paragraph. In the last case, our spectroscopic fit is
meaningless because the spectrum is featureless, but the photometric
fit appears reasonable.  Note that most objects displayed here have
large inferred photometric masses (including the featureless white
dwarf at 1.344 $\msun$), but not all of them. Moreover, we also note
in the case of the magnetic DA+DB double degenerate candidate SDSS
J091016.43$+$210554.20, discussed above and displayed in Appendix
A, that it has the highest inferred photometric
mass in Table \ref{tab2:DD_DADB}, which implies that both magnetic
components of the system are also fairly massive.

\section{Spectral Evolution of DB White Dwarfs}\label{sect2:SpectralEvolution}

In the light of our results, we now focus our attention on the
spectral evolution of DB white dwarfs. In particular, we discuss in
turn the evolution of the DB-to-DA ratio as a function of effective
temperature, and we revisit the question of the origin of hydrogen in
DBA white dwarfs.

\subsection{Evolution of the DB-to-DA Ratio}\label{sect2:DBtoDA}

In order to determine the DB-to-DA ratio as a function of effective
temperature, we retrieved the 27,216 spectra of DA white dwarfs
identified in the SDSS DR7, DR10, and DR12
\citep{DR7,DR10,DR12}. Since we want to characterize the entire
population of DB and DA white dwarfs, we applied no criterion on the
spectral type. Furthermore, to ensure the best possible determination
of the atmospheric parameters for both DB and DA stars, we only kept
the spectra with $\sn > 10$, and the best $\sn$ spectrum for the
objects with multiple observations. The model atmospheres and fitting
technique used to obtain the atmospheric parameters for the DA stars
are described at length in GBB19 and references therein.

Because the SDSS survey is magnitude limited, we need to take into
account that DA and DB white dwarfs with similar $\Te$ and $\logg$
values have different absolute magnitudes (see, e.g., Figure 1 of
\citealt{Bergeron2019}). Therefore, the volume sampled by each white
dwarf type is different. To deal with this issue, we took advantage of
the Gaia trigonometric parallaxes and retained only white dwarfs
within 1 kpc. This left us with a sample of 9863 DA and 1145 DB white
dwarfs. The composition of our sample, subdivided by spectral type, is
presented in Table \ref{tab2:Sample}. We should note that the
atmospheric parameters obtained for the irregular spectral types
(magnetic, composite, etc.) are more uncertain, but this should not
affect significantly our conclusions, since these subtypes represent
only a small fraction of our total sample (see Table
\ref{tab2:Sample}).

\begin{deluxetable*}{ccc||ccc}

\tablecaption{Distribution of the different spectral subtypes in the DA and DB samples \label{tab2:Sample}}
\tablehead{\multicolumn{3}{c}{DA sample} & \multicolumn{3}{c}{DB sample} \\
    \colhead{Spectral Type} & \colhead{Number} & \colhead{Percentage} & \colhead{Spectral Type} & \colhead{Number} & \colhead{Percentage}}
\startdata
	DA & 8440 & 85.57\% & DB\tablenotemark{b} & 859 & 75.02\%\\
	DAH & 236 & 2.39\% & DBH\tablenotemark{c} & 12 & 1.05\% \\
	DAM/DA+M & 725& 7.35\% & DBM/DB+M & 41 & 3.58\%\\
	Other & 99 & 1.00\% & Other & 2 & 0.17\%\\
	Uncertain\tablenotemark{a} & 363 & 3.68\% & Uncertain\tablenotemark{a} & 231 & 20.47\%\\
	Total & 9863 & 100\% & Total & 1145 & 100\% \\
\enddata
\tablenotetext{a}{Includes any spectral type containing ``:''.} 
\tablenotetext{b}{Also includes the DB white dwarfs with traces of hydrogen (DBA) and/or metals (DBZ).}
\tablenotetext{c}{Also includes the DBAH.}
\end{deluxetable*}

One important issue that also needs to be addressed is the
spectroscopic completeness of the SDSS. One class of objects that
was observed in this survey with a high priority is the so-called
``hot standard'' target class, which selects all isolated stars with
clean photometry flags with very blue colors, $u-g<0$ and $g-r<0$,
down to a flux limit of $g<19$ \citep{Eisenstein2006_2}. Most DB white
dwarfs satisfy this condition, as they only become redder below $\Te
\sim 13,000~\K$ or so (see Figure 2 of
\citealt{Bergeron2019}). However, in the case of DA white dwarfs, the
$u-g<0$ criterion is only satisfied for $\Te\ \gta
22,000~\K$. \citet{Eisenstein2006_2} estimated that the completeness
of the SDSS at $u-g>0$ is about 66\% of that at $u-g<0$. Therefore, in
the calculations presented below, we follow the same procedure as that
described in \citet[][see their Section 4]{Eisenstein2006}, and
increase the weight of the stars redder than $u-g=0$ by a factor of
1.5.

\begin{figure}[t]
    \centering
	\includegraphics[clip=true,trim=3cm 9.5cm 3cm 4cm,width=0.8\columnwidth]{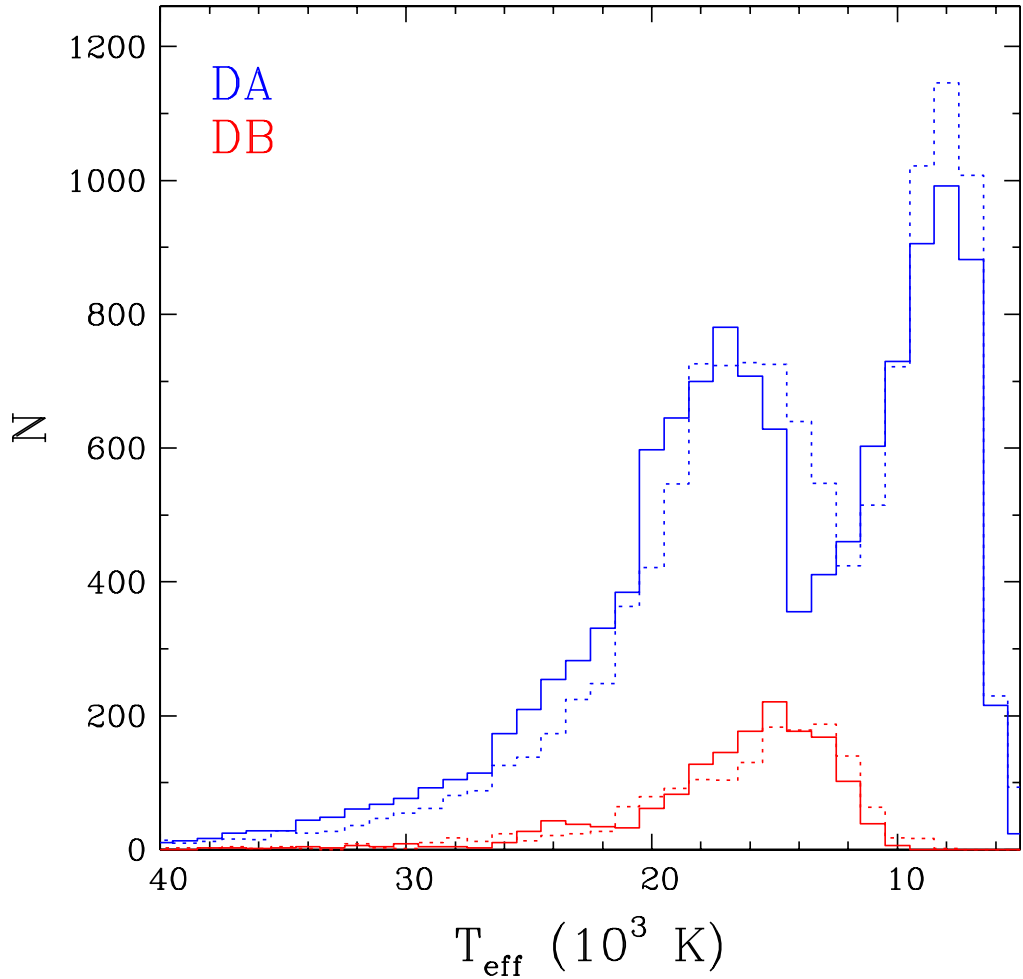}
	\caption{Number of DA and DB white dwarfs in our SDSS sample
          as a function of effective temperature. The solid and dotted
          distributions are based on spectroscopic and photometric
          temperatures, respectively. The objects with $\sn<10$ and
          $D>1~{\rm kpc}$ have been excluded, and a weight factor of
          1.5 has also been applied to the objects with $u-g>0$ (see
          text).}
	\label{fig2:NvsT}
\end{figure}

The distribution of DA and DB white dwarfs as a function of effective
temperature is shown in Figure \ref{fig2:NvsT}, using both
spectroscopic and photometric temperatures. While there are small
differences between the results obtained using these two temperature
scales\footnote{The shift between the photometric and spectroscopic
  distributions of DA stars at high temperatures results from the
  spectroscopic temperatures of DA stars being 10\% higher than the
  photometric values, as discussed in Section \ref{sect2:accuprec} and
  in GBB19.}, the distributions show similar behaviors. In particular,
the number of DB white dwarfs increases monotonically with decreasing
effective temperature, before dropping again at lower temperatures
when DB white dwarfs turn into DC stars, i.e.~when neutral helium
lines become barely detectable ($\Te\,\lesssim 12,000$~K).

More puzzling is the DA distribution, which shows a sudden drop around
$\Te\sim 14,000~\K$ in spectroscopy, and around $\Te\sim 12,000~\K$ in
photometry. The difference in the location of this local minimum can
probably be explained in terms of small inaccuracies in our
spectroscopic temperature scale. Indeed, this corresponds to the
region where the Balmer lines reach their maximum strength, and our
model spectra most likely predict stronger lines than what is actually
observed, causing the spectroscopic solutions to be ``pushed'' on each
side of the maximum (see also Figure 14 of \citealt{Genest2014}). Note
that this may also be caused, in part, by residual calibration issues
with the SDSS spectra, since the same experiment with the DA white
dwarfs from the sample of \citet{Gianninas2011} showed an accumulation
of objects instead of a depletion (see Figure 15 of
\citealt{Genest2014}). But the fact remains that there is a sudden
decrease of DA stars in this temperature range, regardless of the
temperature scale (the increase in the number of DA stars at
even cooler temperatures corresponds to the behavior expected from the white
dwarf luminosity function). It is of course tempting to associate
this drop with the onset of convective mixing\footnote{We remind the
  reader that this process occurs when the bottom of the hydrogen
  convection zone in a DA white dwarf eventually reaches the
  underlying and more massive convective helium envelope, resulting in
  the {\it convective mixing} of the hydrogen and helium layers.} at
low effective temperature ($\Te\,\lesssim 12,000$~K; see Figure 16 of
\citealt{Rolland2018}), which can transform DA white dwarfs into DC
stars (or helium-rich DA stars with a very weak H$\alpha$
feature). But to confirm this hypothesis, one would have to include
all non-DA white dwarfs in this particular range of temperature. Such
a major endeavor is clearly outside the scope of this paper.

\begin{figure}[t]
    \centering
	\includegraphics[clip=true,trim=3cm 9.5cm 3cm 4cm,width=0.8\columnwidth]{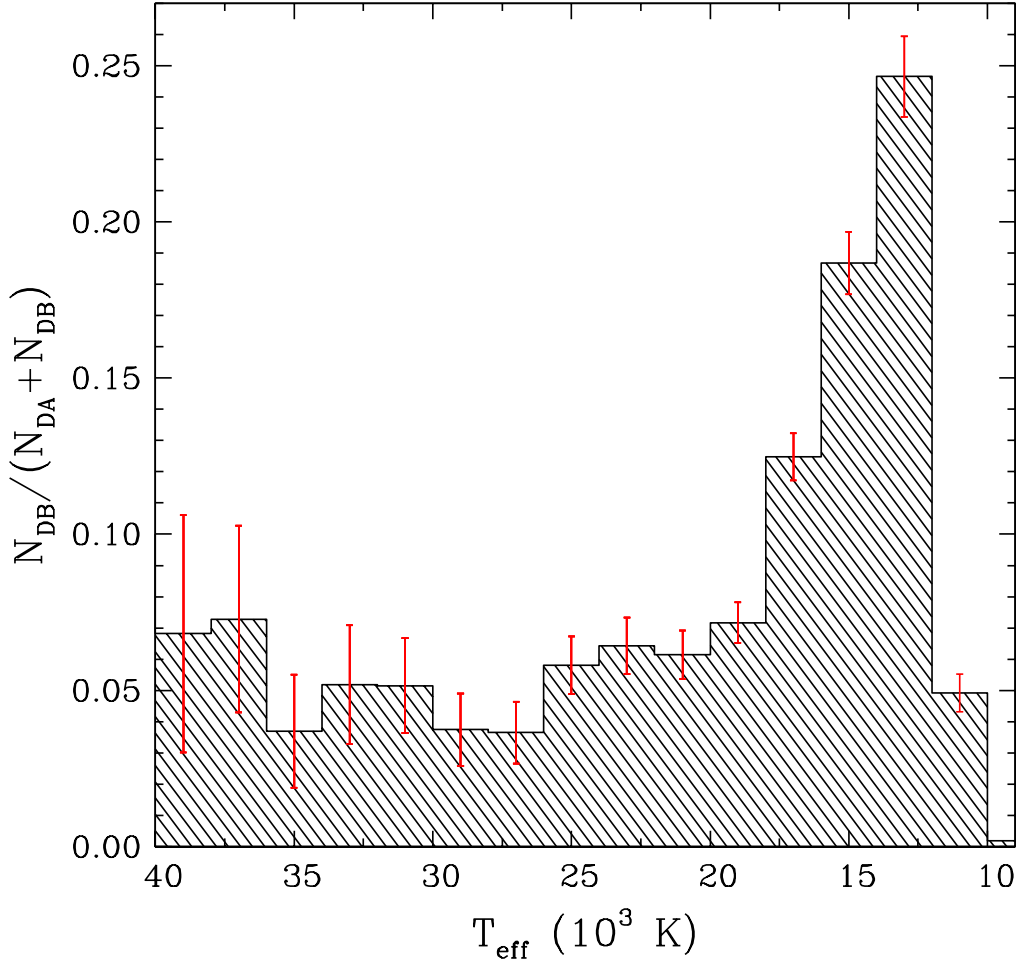}
	\caption{Ratio of the number of DB stars to the total number of DA+DB white dwarfs, as a function of (spectroscopic) effective temperature. The error bars represent the Poisson statistics of each bin. The objects with $\sn<10$ and $D>1~{\rm kpc}$ have been excluded.}
	\label{fig2:evolution}
\end{figure}

Of greater interest in the present context is the fraction of DB white
dwarfs as a function of effective temperature, which we present in
Figure \ref{fig2:evolution}. While we show here the results using
spectroscopic temperatures, the results obtained from photometry are
qualitatively similar. We also remind the reader that the two coolest
temperature bins are not significant because DB white dwarfs turn into
DC stars in this temperature range. Below $\Te=40,000~\K$, down to
$\sim$27,000~K, the DB/(DA+DB) ratio remains fairly constant at
$\sim$5\%, within the uncertainties, and increases slowly to $\sim$7\%
near 20,000~K. Below this temperature, however, the DB/(DA+DB) ratio
rapidly increases to a value of 25\% near 15,000~K\footnote{We note
  that this fraction is only slightly lower when using the photometric
  temperature scale.}, until it drops again at lower temperatures when
DB white dwarfs turn into DC stars. The picture depicted in Figure
\ref{fig2:evolution} is consistent with the results obtained by
\citet{Bergeron2011}, who determined the luminosity functions of DA
and DB stars identified in the PG survey, and found that 20\% of all
white dwarfs below $\Te\sim17,000$~K are DB stars (i.e.~$M_{\rm
  bol}>9.5$ in their Figure 24), while at higher temperatures, only
$\sim$$9$\% of all white dwarfs are DB stars.

The variation of the DB/(DA+DB) ratio observed in Figure
\ref{fig2:evolution} is also entirely consistent with the convective
dilution scenario, where the thin, radiative hydrogen layer present at
the surface of hot DA white dwarfs is being convectively eroded by the
deeper and more massive convective helium envelope, resulting in the
conversion of a DA white dwarf into a DB star. Although detailed
numerical simulations of this convective dilution process are still
unavailable, an examination of the results displayed in Figures 9 and
10 of \citet{Rolland2018} reveals that objects with hydrogen layer
masses in the range $\log M_{\rm H}/\msun=-15.5$ up to $-14$ would
undergo a hydrogen- to helium-atmosphere transition between $\Te =
30,000~\K$ and $18,000~\K$, respectively, in perfect agreement with
the results obtained here (see also \citealt{MacDonald1991}). The fact
that the DB/(DA+DB) ratio increases rather abruptly below 20,000~K
also suggests a narrow range of hydrogen layer masses for the
population of DA stars that undergo the DA-to-DB transition, somewhere
in the order of $\log M_{\rm H}/\msun\sim -14$.

\subsection{Origin of Hydrogen in DBA white dwarfs}\label{sect2:AtmosphericH}

\begin{figure*}[t]
\centering
  \includegraphics[clip=true,trim=0cm 2.6cm 0cm 2.5cm,width=0.85\linewidth]{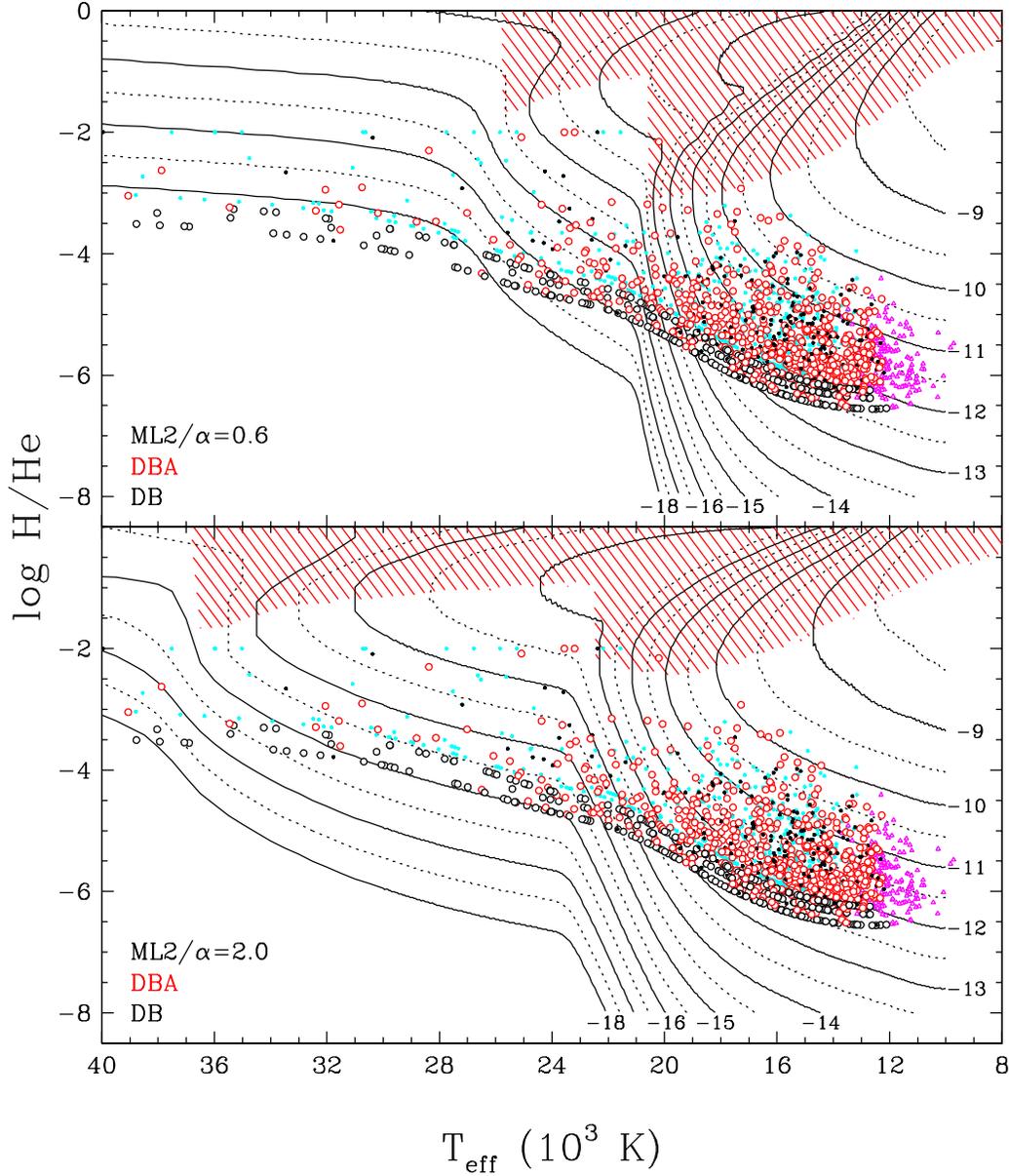}
  \caption{Predicted hydrogen abundances as a function of effective
    temperature (solid and dotted lines) from the simulations of
    \citet{Rolland2018} for homogeneously mixed models at 0.6 $\msun$
    and for both the ML2/$\alpha=0.6$ (upper panel) and $\alpha=2$
    (lower panel) versions of the mixing-length theory. Each curve is
    labeled with the corresponding value of $\log M_{\rm
      H}/M_\odot$. Results from Figure \ref{fig2:h_abundance} are also
    reproduced. The red hatched areas represent the regions through
    which white dwarfs cannot evolve with a constant $\log M_{\rm
      H}/\msun$.}
  \label{fig2:DilutionConvective}
\end{figure*}

Our hydrogen abundance determinations (or limits) from Figure
\ref{fig2:h_abundance} are reproduced in Figure
\ref{fig2:DilutionConvective} together with the detailed simulations
from \citet{Rolland2018}, which show the predictions of the convective
dilution process, where a thin, superficial hydrogen layer of a given
mass has been convectively diluted within the helium envelope,
resulting in a homogeneously mixed H/He convection zone. More
specifically, it is {\it assumed} in these calculations that the
hydrogen layer has been convectively diluted, without paying attention
to dilution process per say. These should thus not be interpreted as
evolutionary sequences in any way. Each curve in Figure
\ref{fig2:DilutionConvective} represents the location of white dwarf
stars with a constant value of $\log M_{\rm H}/M_\odot$, labeled in
the figure. As discussed in Rolland et al., the sudden change of slope
near $\Te\sim20,000$~K corresponds to the temperature where the bottom
of the helium convection zone sinks deep into the white dwarf,
resulting in a further dilution of the photospheric hydrogen within
the deeper helium reservoir as the star cools off.

Also represented as red hatched areas in Figure
\ref{fig2:DilutionConvective} are regions in the $\Te$ -- H/He
parameter space through which white dwarfs cannot evolve continuously
with a constant hydrogen mass (see \citealt{Rolland2018} for a full
discussion). Hence, in order for a white dwarf to cool off with a
constant total mass of hydrogen already homogeneously mixed within the
convective layer, it must be able to evolve {\it continuously} from
the left to the right in this diagram along a single sequence with a
given value of $\log M_{\rm H}/M_\odot$, without crossing this
forbidden region. An examination of the results displayed in Figure
\ref{fig2:DilutionConvective} indicates that the hottest DBA white
dwarfs in our sample ($\Te\gtrsim20,000$~K) can be accounted for by
this scenario if the total hydrogen mass is less than $\log M_{\rm
  H}/M_\odot\sim-15$, assuming the ML2/$\alpha=0.6$ parameterization
of the mixing-length theory (or even less if convection is more
efficient). Note also that in the smallest hydrogen mass models ($\log
M_{\rm H}/M_\odot\lesssim-16$) at high temperatures, there is not
enough hydrogen accumulated at the surface of the star to appear as a
DA white dwarf (see Figures 3 and 4 of \citealt{Manseau2016}). In
other words, the progenitors of some of the hottest DBA white dwarfs
in our sample have never been genuine DA stars, and probably appeared
as stratified DAB stars, with an extremely thin hydrogen layer
floating in diffusive equilibrium at their surface. The hot
($\Te\sim30,000$~K) DAB white dwarf SDSS J1509$-$0108, displayed in
Figure 14 of \citet{Manseau2016}, represents an excellent example of
such a stratified white dwarf, with only $\log M_{\rm
  H}/M_\star=-16.78$ ($\log M_{\rm H}/M_\odot=\log M_{\rm
  H}/M_\star+0.22$ for a $M_\star\sim0.6\ \msun$ white dwarf).
  
The other hot ($\Te>20,000$~K) DB white dwarfs in Figure
\ref{fig2:DilutionConvective} with no detectable traces of hydrogen
have so little inferred total hydrogen mass in their stellar envelope
that these stars always appeared as DB white dwarfs, and their
immediate progenitors are most certainly the hot DB stars in the
DB-gap analyzed by \citet{Eisenstein2006}. These obviously will remain
DB stars, with no detectable traces of hydrogen, throughout their
evolution. More importantly, even the hot DBA stars in our sample
above $\Te\sim 20,000$~K await a similar fate, given that the
deepening of the mixed H/He convection zone at lower temperatures will
completely dilute any residual hydrogen left in the stellar envelope,
well below the limit of visibility of hydrogen, H$\alpha$ in this
case. In this respect, we disagree with the conclusions of
\citet{Koester2015} who suggested that practically all DB white dwarfs
probably show some trace of hydrogen if the spectroscopic resolution
and S/N are high enough. Regardless of the observational limit, there
is obviously also a {\it theoretical limit} on H/He, predicted by
model atmospheres, below which hydrogen becomes invisible (see in
particular Figure 3 of \citealt{Rolland2018}). Such threshold limits
are certainly achieved according to the results displayed in Figure
\ref{fig2:DilutionConvective}. Hence we conclude that, {\it according
  to the convective dilution scenario alone}, all hot ($\Te>20,000$~K)
DB and DBA white dwarfs will most likely evolve into nearly pure
helium-atmosphere white dwarfs at lower temperatures, with no
detectable traces of hydrogen. In contrast, DBA stars at lower
temperatures should retain a trace of hydrogen much longer since the
depth of the mixed H/He convection zone remains almost constant below
20,000~K (see Figure 9 of \citealt{Rolland2018}), thus reducing any
further dilution of hydrogen.

We now turn our attention to the bulk of DBA white dwarfs in our
sample, found below $\Te\sim20,000$~K in Figure
\ref{fig2:DilutionConvective}. In order to account for the observed
photospheric hydrogen abundances in these stars, the total hydrogen
mass present in the stellar envelope must be in the range $-14<\log
M_{\rm H}/M_\odot<-10$, regardless of the convective efficiency. The
problem here is that hotter DA progenitors with such thick hydrogen
layers in diffusive equilibrium at their surface --- i.e.~with
chemically stratified envelopes --- would inhibit convection in the
deeper helium envelope (see Figure 11 of \citealt{Rolland2018}),
preventing any DA-to-DB conversion in this temperature range. In DA
white dwarfs with such thick hydrogen layers, mixing between the {\it
  convective hydrogen layer} and the deeper, and more massive helium
envelope would eventually occur, but at much lower effective
temperatures ($\Te\ \lesssim 12,000$~K) according to the calculations
of \citet[][see their Figure 16]{Rolland2018}. Hence we must conclude
that the total amount of hydrogen present in the bulk of DBA white
dwarfs below $\Te\sim20,000$~K is too large to have a residual origin
resulting from the convective dilution scenario, and that other
sources of hydrogen must be invoked, such as accretion from the
interstellar medium, comets, or disrupted asteroids, etc., a
conclusion also reached in several previous investigations (e.g.,
\citealt{MacDonald1991}, \citealt{Bergeron2011},
\citealt{Koester2015}, \citealt{Rolland2018}). Despite these
discrepancies, the convective dilution scenario remains the only
viable explanation for the transformation of DA into DB white dwarfs
below 20,000~K.

Finally, we note that all the DBA white dwarfs in our sample
completely avoid the so-called forbidden region in Figure
\ref{fig2:DilutionConvective} (defined by the red hatched areas),
especially with the ML2/$\alpha=0.6$ models. Note, in particular, how
the most hydrogen-rich DBA stars in our sample (including PG
1311+129), already discussed in Section \ref{sect2:DBA}, follow nicely
the bottom limit of the forbidden region. Because the convective
dilution process, which transforms a chemically stratified DA star
into a homogeneous DB white dwarf, makes somehow the object ``jump''
over this forbidden region, the extreme DBA stars in our sample are
probably in the process of being convectively mixed, with hydrogen
constantly trying to float back to the surface, as suggested by
\citet{Bergeron2011} in the case of PG 1311+129, which incidentally,
shows small spectroscopic variations as a function of time (see their
Figure 28).

\section{Discussion}\label{sect2:conclusion}

We presented a detailed model atmosphere analysis of DB white dwarfs
drawn from the SDSS database. The large set of optical spectra and
$ugriz$ photometry available from the SDSS, coupled with the exquisite
trigonometric parallax measurements from the {\it Gaia} mission,
allowed us to measure the fundamental parameters of DB white dwarfs
using both the spectroscopic and photometric techniques. In turn, the
results obtained from both fitting techniques provided us with a
unique opportunity to assess the precision, but more importantly the
{\it accuracy} of the measurements, in particular the effective
temperature, stellar mass, and photospheric composition, which form
the basic elements for understanding the origin and evolution of these
stars.

We identified several problems in our analysis, both theoretical and
observational. On the theoretical front, van der Waals broadening
remains the largest source of uncertainty in the calculations of
synthetic spectra for DB white dwarfs. Nevertheless, our mass
comparison displayed in Figure \ref{fig2:correltm} reveals that the
spectroscopic masses at low temperatures appear fairly reasonable, if
we exclude the objects with extremely weak helium lines. We also see
some clear evidence for 3D hydrodynamical effects in our spectroscopic
mass distributions in the $16,000-22,000$~K temperature range, and
additional work along the lines of \citet{Cukanovaite2018}, but by
including traces of hydrogen, should improve the spectroscopic mass
measurements significantly. Finally, given the large spectroscopic
data set available here, it would probably be worth revisiting the
question of the convective efficiency in hot DB stars --- which in all
recent investigations, including ours, has been set to
ML2/$\alpha=1.25$ --- by repeating the experiments performed in
Section 4.4 of \citet{Bergeron2011}.

We identified a few problems on the observational front as well. While
the number of white dwarfs discovered by the SDSS has been increased
by more than a factor of 10 since the latest version of the Villanova
White Dwarf Catalog of \citet{McCookSion1999}, the optical spectra
from the SDSS still appear to suffer from some residual flux
calibration problems. This becomes apparent at high effective
temperatures, where the spectroscopic masses are lower than those
measured at cooler temperatures, or lower than those inferred from
photometry for that matter. This problem is observed for both DA and
DB stars. Furthermore, even though the large number of white dwarf
spectra available in the SDSS has significantly improved our
understanding of these objects, the generally low S/N of the SDSS
spectra (see Figure \ref{fig2:SN}) still represents a strong
limitation in terms of making more precise spectroscopic measurements
of white dwarf parameters, and it is hoped that future spectroscopic
surveys will improve the situation.

Our analysis has uncovered several interesting objects, including the
existence of massive DB white dwarfs, mostly below $\Te\sim18,000$~K,
whose progenitor might be the Hot DQ stars. We also identified a large
population of unresolved double degenerate binaries composed of two DB
white dwarfs, and even occasionally DB+DA double degenerates. As in
previous investigations, we did not find any evidence for single
low-mass ($M<0.48\ \msun$) DB white dwarfs, obviously because nature
does not produce such objects. All low-mass white dwarfs appear to
have hydrogen-rich atmospheres. As discussed by \citet{Bergeron2011},
common envelope evolution is required to produce such low mass white
dwarfs, and this close-binary evolutionary channel must produce white
dwarfs with hydrogen layers too massive to allow the DA to DB
conversion at any temperature.

Probably the most important key result of our analysis is the hydrogen
abundance (or limits) distribution as a function of effective
temperature, summarized in Figure \ref{fig2:DilutionConvective}. This
picture represents a static view, or a snapshot in time, of the
current distribution of hydrogen abundances in DB and DBA white
dwarfs, which detailed evolutionary calculations must eventually
explain. In particular, DA white dwarfs must be able to turn into DB
stars in the appropriate temperature range, and properly account for
the observed DB-to-DA ratio. The residual amount of hydrogen expected
from the DA-to-DB transition is much smaller than that inferred from
current homogeneous envelope models \citep{Rolland2018}, unless the
hydrogen-to-helium abundance ratio measured using homogeneous model
atmospheres is somehow overestimated, for instance, if the atmosphere
is chemically inhomogeneous (\citealt{MacDonald1991},
\citealt{Genest2017}). The alternative is to have an external source
of hydrogen --- either from the interstellar medium, disrupted
asteroids, small planets, or even comets --- polluting the atmosphere
of DB white dwarfs after the DA-to-DB transition has occurred.

Yet, another alternative proposed by B. Rolland et al.~(2019, in
preparation) is to have an {\it internal} source of hydrogen. With
this new paradigm, hydrogen is initially diluted in the deep stellar
envelope of a hot PG1159 progenitor. With time, hydrogen slowly
diffuses upward, slowly building a superficial hydrogen-rich
atmosphere. However, calculations show that when the convective
dilution process occurs, the resulting abundance profile is far from
equilibrium, and large amounts of hydrogen are still present in the
deep interior. As the convection zone of the cooling DB stars grows,
{\it large amounts of hydrogen is being dredged-up to the surface}, a
phenomenon similar to that invoked in the context of DQ white dwarfs
\citep{Pelletier1986}.

Also of importance in this context is the existence, and persistence,
of pure DB stars, i.e.~helium-atmosphere white dwarfs that show no
hydrogen feature whatsoever. As discussed in our analysis, this is not
a simple problem of detectability, because envelope models such as
those displayed in Figure \ref{fig2:DilutionConvective} do predict
hydrogen abundances where H$\alpha$ simply becomes {\it undetectable},
even theoretically. White dwarfs with such small total amounts of
hydrogen present in their envelope --- in particular those discovered
in the DB gap --- can only be explained by the so-called born-again
post-AGB (asymptotic giant branch) evolutionary scenario, involving a
very late helium-shell flash, or an AGB final thermal pulse, as
described in detail by \citet{Werner2006}. As discussed in
\citet{Bergeron2019}, it is possible that in this process, the total
mass of helium could also be significantly reduced. This could perhaps
explain the origin of cool DQ white dwarfs, in which carbon is being
dredged-up from the core, a process that occurs only if the helium
envelope is not too thick according to \citet{Pelletier1986}. If this
interpretation is correct, the immediate progenitors of DQ stars could
be the ``pure'' DB white dwarfs discussed above. By the same token,
one could imagine that these very late helium-shell flashes are
violent enough as to wipe out the stellar environment from any
planetary debris, thus providing a natural explanation for the absence
of metals in cool DQ stars.

Clearly, all these issues will require larger and better data sets ---
higher S/N spectra in particular --- but most importantly,
state-of-the-art evolutionary model calculations able to predict the
spectral evolution of white dwarf stars as a function of their cooling
age. Then perhaps one day, we will be able to map the range of total
hydrogen mass ($M_{\rm H}$) present in these stars --- arguably the
Holy Grail of white dwarf astrophysics.

We are grateful to Patrick Tremblay for useful discussions regarding
the behavior of DB model spectra. This work is supported in part by
the NSERC Canada and by the Fund FRQ-NT (Qu\'ebec).

\clearpage
\appendix
\section{DA+DB unresolved double degenerate systems}\label{annexe_B}

\begin{figure}[h]
\centering
        \includegraphics[clip=true,trim=1.5cm 0.5cm 1.5cm 0.5cm,width=0.7\linewidth]{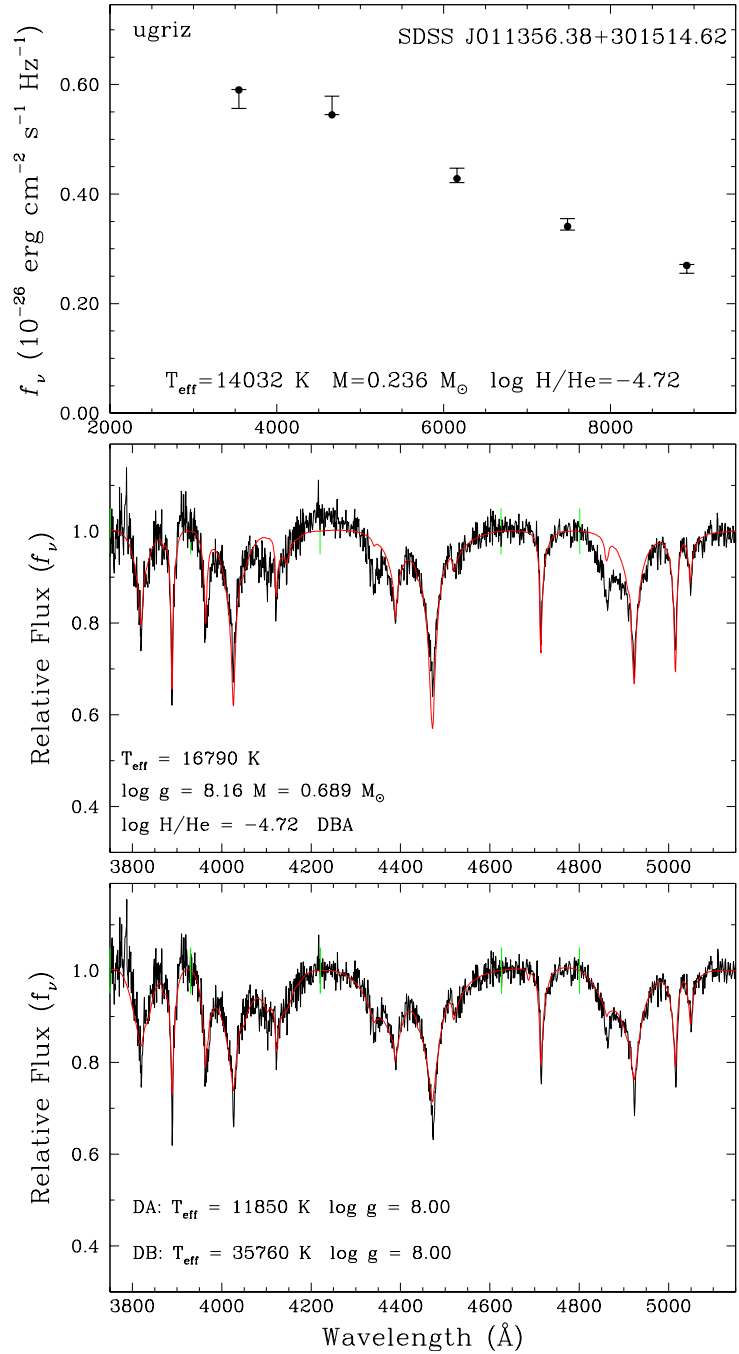}
        \caption{SDSS J011356.38+301514.62. The complete figure set (10 images) is available in the online journal.}
\end{figure}

\newpage

\bibliographystyle{aasjournal}
\bibliography{references}

\end{document}